\pdfoutput=1

\documentclass[11pt]{article}

\usepackage[preprint]{acl}

\usepackage{times}
\usepackage{latexsym}
\usepackage{cleveref} 
\crefname{figure}{Fig.}{Figs.}  
\crefname{table}{Table}{Tables}  

\usepackage[T1]{fontenc}

\usepackage{array}
\usepackage{amssymb}
\usepackage{pifont}
\usepackage{booktabs}
\usepackage{xspace}
\usepackage{arydshln}
\usepackage{multirow}
\usepackage{changes}
\usepackage{changepage} 

\usepackage{xcolor}
\definecolor{darkgreen}{rgb}{0.0, 0.5, 0.0}
\definechangesauthor[name={A}, color=blue]{A}
\definechangesauthor[name={N}, color=red]{N}
\definechangesauthor[name={R}, color=darkgreen]{R}
\definechangesauthor[name={L}, color=brown]{L}
\definechangesauthor[name={O}, color=orange]{O}
\definechangesauthor[name={E}, color=pink]{E}

\definecolor{DarkGreen}{rgb}{0.0, 0.5, 0.0} 
\definecolor{DarkYellow}{rgb}{0.8, 0.6, 0.0} 
 \definecolor{DarkPurple}{rgb}{0.5, 0.0, 0.5}

\usepackage[utf8]{inputenc}

\usepackage{microtype}

\usepackage{inconsolata}
\usepackage{enumitem} 

\usepackage{graphicx}
\usepackage{svg}
\renewcommand{\footnotesize}{\scriptsize}

\title{REAL-MM-RAG: A Real-World Multi-Modal Retrieval Benchmark}

\author{
    Navve Wasserman\textsuperscript{1,2}, Roi Pony\textsuperscript{1}, Oshri Naparstek\textsuperscript{1}, 
    Adi Raz Goldfarb\textsuperscript{1}, Eli Schwartz\textsuperscript{1} \\
    \textbf{Udi Barzelay\textsuperscript{1}, Leonid Karlinsky\textsuperscript{1}} \\
    \textsuperscript{1}IBM Research Israel \quad \textsuperscript{2}Weizmann Institute of Science
}

\begin{document}
\maketitle

\begin{abstract}
Accurate multi-modal document retrieval is crucial for Retrieval-Augmented Generation (RAG), yet existing benchmarks do not fully capture real-world challenges with their current design. We introduce REAL-MM-RAG, an automatically generated benchmark designed to address four key properties essential for real-world retrieval: (i) multi-modal documents, (ii) enhanced difficulty, (iii) Realistic-RAG queries and (iv) accurate labeling.
Additionally, we propose a multi-difficulty-level scheme based on query rephrasing to evaluate models' semantic understanding beyond keyword matching.
Our benchmark reveals significant model weaknesses, particularly in handling table-heavy documents and robustness to query rephrasing. To mitigate these shortcomings, we curate a rephrased training set and introduce a new finance-focused, table-heavy dataset. Fine-tuning on these datasets enables models to achieve state-of-the-art retrieval performance on REAL-MM-RAG benchmark. Our work offers a better way to evaluate and improve retrieval in multi-modal RAG systems while also providing training data and models that address current limitations.
\end{abstract}

\section{Introduction}
\label{sec:intro}

\begin{figure*}[ht!]
    \centering
    \includegraphics[width=1\textwidth]{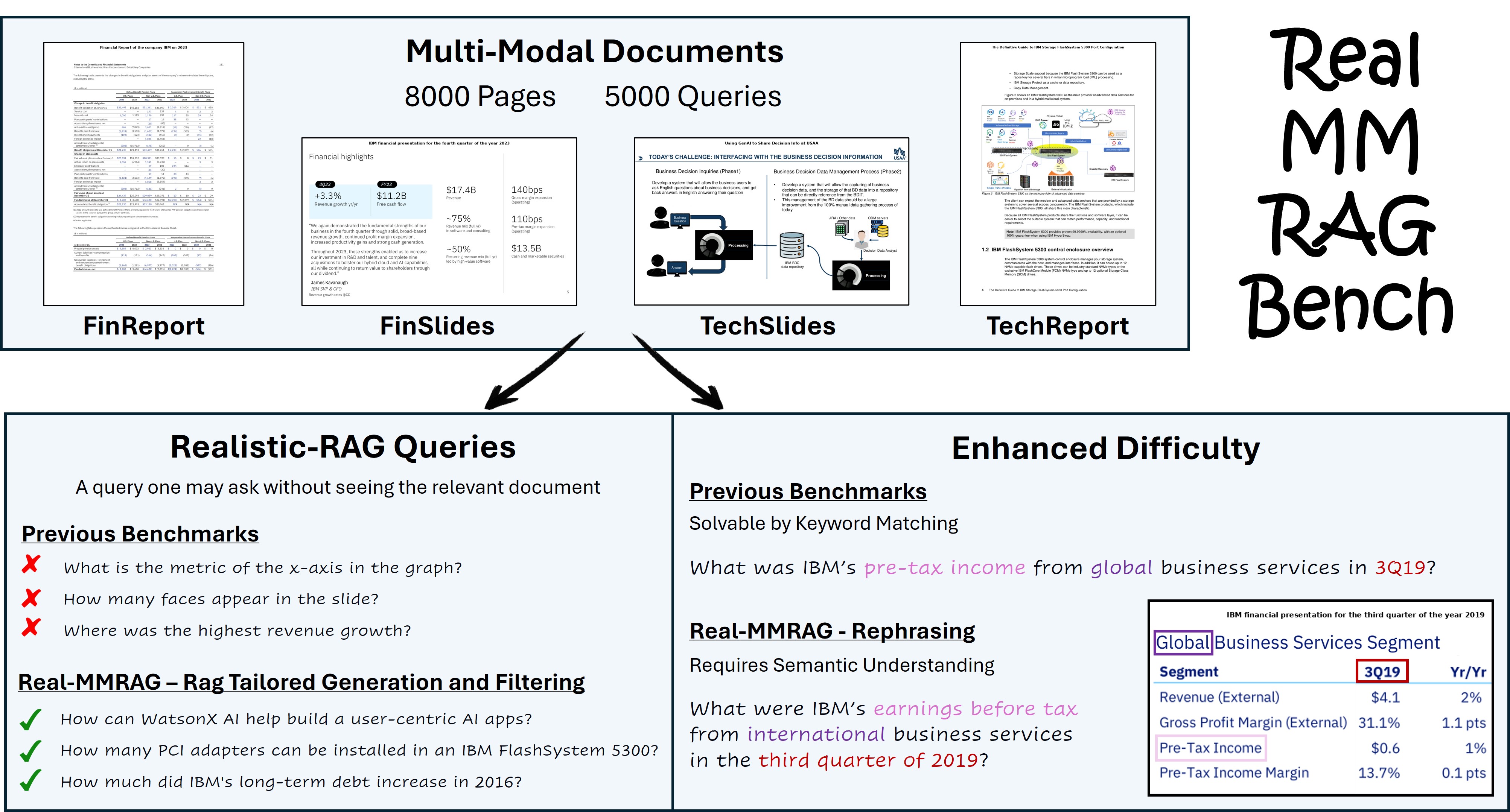} 
    \caption{\textbf{Proposed Real-MM-RAG Benchmark}}
    \vspace{-0.2cm}
    \label{fig:Benchmark} 
\end{figure*}

Accurate retrieval of relevant documents is a cornerstone of modern natural language processing (NLP) applications, whether used alone or in advanced pipelines like Retrieval-Augmented Generation (RAG). RAG~\citep{lewis2020retrieval} has emerged as a powerful approach wherein models retrieve external information before generating answers or content, enabling operation over large document collections. \emph{Multi-modal RAG} extends this to real-world scenarios involving text, figures, tables, and potentially entire page images.\\

\vspace{-0.15cm}
\noindent
Successful retrieval is crucial for RAG, as retrieving the wrong page or document inevitably hinders the final generated response. Therefore our analysis will focus on the retrieval part. Although research on RAG is advancing, the field still lacks a complete understanding of how models perform in realistic setups, both for evaluating performance and identifying current weaknesses to overcome. This gap arises from a shortage of benchmarks that thoroughly assess real-world retrieval challenges. \\

\vspace{-0.15cm}
\noindent
We identify four essential properties for a real-world document retrieval benchmark, particularly in multi-modal contexts: \textbf{\emph{(i) Multi-modal documents:}} The dataset should include pages with text, figures, tables, and other visual elements to reflect the complexity of real-world materials.
\textbf{\emph{(ii) Enhanced difficulty:}} Queries should require more than simple keyword matching and involve a large corpus of contextually similar pages to ensure challenging evaluations. \textbf{\emph{(iii) Realistic-RAG queries:}} Questions must be posed naturally, without explicit references to pages—reflecting queries a person might ask when seeking information without knowing the answer's location (in contrast to generic question-answering setups).
\textbf{\emph{(iv) Accurate labeling:}} All documents relevant to a query must be correctly and exhaustively labeled to prevent underestimation of retrieval performance and to avoid false negatives. \\

\vspace{0.1cm}
\noindent
Although a few recent benchmarks touch on some of these aspects~\citep{faysse2024colpali,ma2024unifying,ma2024mmlongbench}, most fail to fully capture them, limiting their usefulness for understanding and improving multi-modal retrieval models. We introduce \emph{\textbf{REAL-MM-RAG-Bench}} \emph{(Real-World Multi-Modal Retrieval-Augmented Generation Benchmark)}, a benchmark designed to satisfy the properties above:

\emph{\textbf{Multi-modal documents (Property i):}} Our dataset comprises slides and documents with text, figures, tables, and images, requiring systems to handle combined textual and visual data.

\emph{\textbf{Enhanced difficulty (Property ii):}} Instead of relying on isolated pages or trivial queries, we focus on long documents from specialized domains (e.g., IBM finance reports, FlashSystem technical materials). 
This makes retrieval significantly more challenging, as models must differentiate between highly similar content within the same domain. Additionally, we incorporate a rephrasing step to ensure that query wording and order are not identical to the page content, requiring semantic understanding rather than simple keyword matching.

\emph{\textbf{Realistic-RAG queries (Property iii):}} We use a two-step process to generate queries: vision-language models (VLMs) create retrieval-focused queries, and large language models (LLMs) filter them to ensure natural phrasing and realistic user intent. Unlike many existing datasets, our queries avoid direct references to specific pages, reflecting authentic retrieval scenarios. 

\emph{\textbf{Accurate labeling (Property iv):}} Ensuring all pages answering a query are correctly identified is crucial, especially in benchmarks with many similar pages. Existing benchmarks often mislabel valid documents as incorrect, leading to false negatives. To address this, we employ an automated pipeline using VLMs to verify query relevance to each page. While computationally intensive, this approach enhances labeling reliability, particularly for closely related pages.

\vspace{0.12cm}
\noindent
Our benchmark enables reliable evaluation of current retrieval models and uncovering some of their weaknesses. Additionally, it incorporates two essential properties that expose specific retrieval challenges: 

\emph{\textbf{Rephrasing Robustness Evaluation:}} In real-world scenarios, users rarely phrase their queries exactly as they appear in documents. However, both VLMs and human annotators tend to generate queries that closely mirror the source material, often using similar words and sentence structures~\citep{smeaton1998user,zhu2024enhancing}. This fails to reflect natural user behavior, where users are not directly exposed to the document page when forming queries.
To address this, we introduce a multi-level rephrasing benchmark, modifying queries at three distinct levels—ranging from slight rewording to significant structural changes. Our experiments show that current retrieval models struggle to maintain performance across these variations, highlighting a critical weakness in their semantic understanding.

\emph{\textbf{Table-Focused Scenarios:}} Table-heavy documents (e.g. financial reports) often contain dense tabular data, posing a major challenge for retrieval models. By incorporating table-heavy documents into our benchmark, we expose key deficiencies in table comprehension that significantly impact model performance.
These properties allow us to demonstrate that all current retrieval models exhibit weaknesses in handling both rephrased queries and table-heavy financial documents. 

\noindent
To address these shortcomings, we leverage insights from our benchmark to enhance retrieval performance. Specifically, we introduce two targeted training strategies: (i) a \emph{rephrased training dataset}, generated by rephrasing the ColPali training dataset~\citep{faysse2024colpali}, and (ii) a \emph{finance-table-heavy training set}, designed to improve retrieval in tabular contexts. Fine-tuning the current best model on these datasets achieves state-of-the-art retrieval performance on our benchmarks. This demonstrates how systematic evaluation through our benchmark can informs effective training strategies, leading to more robust and adaptable retrieval models.

\vspace{0.23cm}
\noindent
\textbf{The contributions of this paper are as follows:}
\vspace{-0.2cm}
\begin{itemize}[leftmargin=*] 
    \setlength{\itemsep}{3pt}
    \setlength{\parskip}{0pt}
    \setlength{\itemindent}{1pt} 
    \item Defining properties of a real-world retrieval benchmark and highlighting shortcomings in existing ones.
    \item Introducing a high-quality multi-modal retrieval benchmark with an automated pipeline for query generation, filtering, and labeling verification.
    \item Establishing a rephrasing robustness evaluation framework via multi-level query rephrasing.
    \item Providing two specialized training datasets: (i) a rephrased dataset and (ii) a finance-table-heavy dataset, where fine-tuning on them significantly enhances retrieval performance.
\end{itemize}

\section{Related Work}
\label{sec:related_work}

\subsection{Text-Based Retrieval}  

Text-based retrieval methods identify relevant documents given a query and are widely used in RAG systems. Lexical matching techniques like BM25~\cite{robertson1994okapi} and TF-IDF~\cite{sparck1972statistical} are efficient but lack semantic understanding. Sparse models like SPLADE~\cite{formal2021splade} improve retrieval by expanding queries into high-dimensional sparse representations but struggle with deep contextual meaning. Dense retrieval models, leveraging transformers like BERT~\cite{devlin2018bert}, T5~\cite{raffel2020exploring}, and DPR~\cite{karpukhin2020dense}, map queries and documents into a continuous vector space, enhancing recall but demanding significant computational resources for training and inference. Hybrid methods, that combine lexical and dense retrieval, such as ColBERT~\cite{khattab2020colbert} and ANCE~\cite{xiong2020approximate}, often achieving state-of-the-art performance. A recent model, M3-Embedding~\cite{chen2024bge}, unifies dense, sparse, and multi-vector embeddings, achieving strong retrieval performance. Despite advancements, text-based retrieval struggles with multi-modal content, particularly in scenarios where visual cues enhance contextual understanding.

\subsection{Multi-Modal Retrieval}

Until recently, multi-modal retrieval primarily relied on Optical Character Recognition (OCR) to extract textual information from documents, including text within visual elements. More recent approaches detect visual components and process them in one of two ways: (i) Captioning-based retrieval, where a VLM generates textual descriptions of visual elements, enabling standard text-based retrieval~\citep{ramos2023retrieval}; or (ii) Direct embedding, where visual elements are embedded either using VLMs directly or through contrastive Vision-Language Models that align separate visual and text encoders via contrastive losses~\citep{radford2021learning,zhai2023sigmoid}.

\vspace{0.1cm}
\noindent
A more recent line of work leverages the strong performance of VLMs in analyzing full document images by embedding entire document pages instead of relying on OCR-based extraction. 
Methods such as VISRAG~\citep{yu2024visrag} and DSE~\citep{ma2024unifying} generate dense embeddings directly from document images. Similarly, ColPali~\citep{faysse2024colpali} generates multi-vector embeddings for ColBERT-style late interaction retrieval, using PaliGemma~\citep{beyer2024paligemma} or in a newer variant, ColQwen, utilizes Qwen2-VL~\citep{Qwen2VL}.
These methods have demonstrated significant improvements over earlier text-based and OCR-dependent retrieval approaches. Our benchmark provides a rigorous evaluation framework for both text-based and visual-based multi-modal retrieval.

\begin{figure*}[t]
    \centering
    \includegraphics[width=1\textwidth]{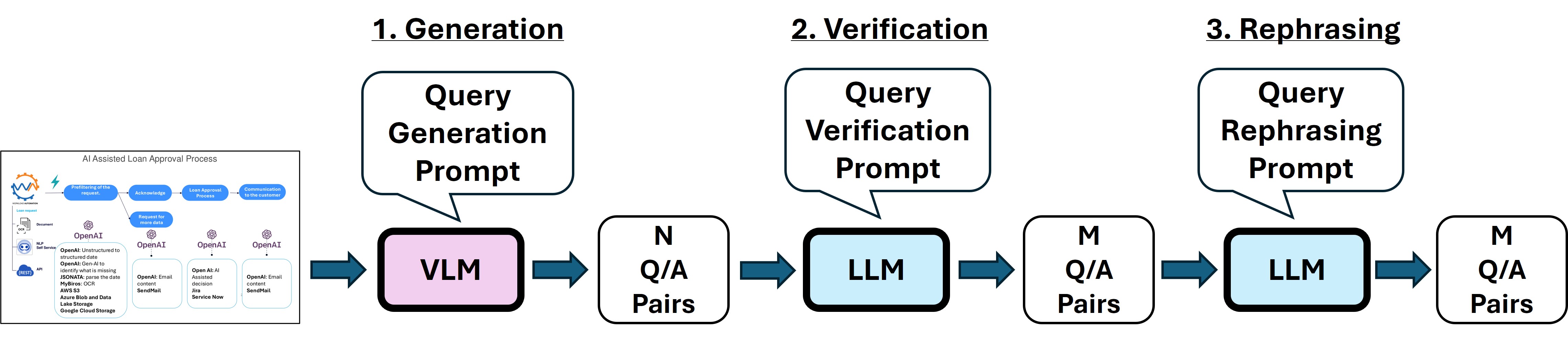} 
    \caption{\textbf{Benchmark Construction Pipeline}}
    \label{fig:Data_generation_pipeline} 
\end{figure*}

\subsection{Query Rephrasing}

Retrieval models known to be highly vulnerable to query rephrasing~\citep{zuccon2016query,bailey2017retrieval,sidiropoulos2022analysing,penha2022evaluating,hagen2024revisiting}, often leading to significant performance degradation. However, only few works have provided accessible and reliable evaluation frameworks for model robustness. An early study~\citep{bailey2016uqv100} introduced a basic Query Variability dataset, while more recent works~\citep{benham2018towards,lu2019relevance,penha2022evaluating} focus on automatically generating query variations. Yet, no prior research has established a standardized benchmark for retrieval robustness, nor a dedicated RAG robustness benchmark—especially for multi-modal retrieval. In contrast, we leverage LLMs to generate multi-level query rephrasing, enabling structured comparative evaluation for multi-modal documents retrieval.

\subsection{Multi-Modal Retrieval Benchmarks}

Despite the growing importance of document retrieval in multi-modal RAG systems, only a few evaluation benchmarks exist, all of which fall short in key aspects crucial for real-world scenarios. We review recent efforts and highlight their limitations (with further comparison in \cref{Table:benchmark_comparison} and \cref{sec:Benchmarks_Comparisons}).
While many question-answering benchmarks exist~\citep{mathew2021docvqa,zhu2022towards,masry2022chartqa,islam2023financebench,ding2024mvqa}, they are largely unsuitable for RAG. Their queries assume exposure to a specific page (unlike RAG), and their tendency toward high variability make retrieval easier. Some benchmarks, such as MMLongBench~\citep{ma2024mmlongbench} (based on 130 lengthy PDFs) and SlideVQA~\citep{tanaka2023slidevqa}, are partially relevant but not suited for RAG (see \cref{sec:Benchmarks_Comparisons}).

\vspace{0.1cm}
\noindent
A notable benchmark is WIKI-SS-NQ~\citep{ma2024unifying}, generated from Wikipedia screenshots with real human queries—the only dataset providing mostly valid retrieval queries. However, it is not multi-modal benchmark, consists of mainly text-based documents, and has narrow sub-domain coverage.
The ViDoRe benchmark~\citep{faysse2024colpali}, introduced in the ColPali paper, comprises both QA datasets and domain-specific documents with generated queries filtered by human annotators. While the QA datasets are unsuitable for RAG, the domain-specific queries are better tailored but suffer from trivial difficulty (e.g., synthetic datasets reporting an NDCG@5 >~95). This occurs because pages often differ significantly, and VLM-generated questions closely mirror the original page wording, making retrieval easy.

\vspace{0.1cm}
\noindent
Our REAL-MM-RAG-Bench is the first multi-modal retrieval benchmark incorporating all essential properties for real-world RAG. It features a challenging setup with broad sub-domain coverage, long documents, RAG-tailored rephrased queries, and accurate labeling. Additionally, it is the first to offer a robustness evaluation through multi-level query rephrasing.

\section{REAL-MM-RAG-Bench}
\label{sec:benchmark}

\begin{table*}[t]
\footnotesize
\renewcommand{\arraystretch}{1.5} 
\vspace{-0.15cm}
\centering
\begin{tabular*}{0.89\linewidth}{lccccccccc} 
\toprule
{} 
& \multicolumn{2}{c}{\textbf{Statistics}} 
& \multicolumn{1}{c}{\textbf{Multi-Modal}} 
& \multicolumn{3}{c}{\textbf{Enhanced Difficulty}} 
& \multicolumn{2}{c}{\textbf{Realistic-RAG Queries}} 
& {\textbf{Accurate Labels}} 
\\
\midrule

{} 
& {\textbf{$\#$ }} 
& {\textbf{$\#$ }} 
& {\textbf{MM}} 
& {\textbf{Long}} 
& {\textbf{Sub}} 
& {\textbf{Queries}} 
& {\textbf{RAG}} 
& {\textbf{RAG}} 
& {\textbf{False }} 
\\

{} 
& {\textbf{Pages}} 
& {\textbf{Queries}} 
& {\textbf{Pages}} 
& {\textbf{Docs}}
& {\textbf{domain}} 
& {\textbf{Rephr-}} 
& {\textbf{Tailored}} 
& {\textbf{Query}} 
& {\textbf{Neg.}} 
\\

{\textbf{Benchmark}} 
& {} 
& {} 
& {\textbf{}} 
& {\textbf{}} 
& {\textbf{Cover}} 
& {\textbf{asing}} 
& {\textbf{Gen.}} 
& {\textbf{Verif.}} 
& {\textbf{Verif.}} 
\\

\midrule
SlideVQA  & 52k & 14.5k & \textcolor{DarkGreen}{\ding{51}}  & \textcolor{DarkGreen}{\ding{51}} & \textcolor{red}{\ding{55}} &  \textcolor{red}{\ding{55}} &  \textcolor{red}{\ding{55}} &  \textcolor{red}{\ding{55}} &  \textcolor{red}{\ding{55}} \\ 
MMLONG & 7k & 1k & \textcolor{DarkGreen}{\ding{51}} &  \textcolor{DarkGreen}{\ding{51}}  &  \textcolor{red}{\ding{55}} &  \textcolor{red}{\ding{55}} &  \textcolor{red}{\ding{55}} &  \textcolor{red}{\ding{55}} &  \textcolor{red}{\ding{55}}\\
WIKI-SS-NQ  & 4k & 4k & \textcolor{red}{\ding{55}} &  \textcolor{red}{\ding{55}}  &  \textcolor{red}{\ding{55}} &  \textcolor{red}{\ding{55}} &  \textcolor{DarkGreen}{\ding{51}} &  \textcolor{DarkYellow}{\ding{51}\raisebox{-0.9ex}{\textsuperscript{\kern-0.9em\scalebox{1.6}{\ding{55}}}}} &  \textcolor{red}{\ding{55}}\\
ViDoRe  & 8k & 4k & \textcolor{DarkGreen}{\ding{51}} &  \textcolor{red}{\ding{55}} & \textcolor{red}{\ding{55}} &   \textcolor{red}{\ding{55}} &  \textcolor{DarkYellow}{\ding{51}\raisebox{-0.9ex}{\textsuperscript{\kern-0.9em\scalebox{1.6}{\ding{55}}}}}&  \textcolor{DarkYellow}{\ding{51}\raisebox{-0.9ex}{\textsuperscript{\kern-0.9em\scalebox{1.6}{\ding{55}}}}} &  \textcolor{red}{\ding{55}}\\
\midrule
Ours  & 
8k & 5k & \textcolor{DarkGreen}{\ding{51}} &  \textcolor{DarkGreen}{\ding{51}} & \textcolor{DarkGreen}{\ding{51}} &  \textcolor{DarkGreen}{\ding{51}}  &  \textcolor{DarkGreen}{\ding{51}} &  \textcolor{DarkGreen}{\ding{51}} &  \textcolor{DarkGreen}{\ding{51}}\\
\bottomrule 
\end{tabular*}
\caption{
\textbf{Document Retrieval Benchmarks Comparison.}}
\label{Table:benchmark_comparison}
\vspace{-0.22cm}
\end{table*}

Creating a high-quality benchmark manually is both exhaustive and error-prone, limiting its size and reliability. To address this, we propose an \textbf{\emph{automated generation and verification pipeline}} tailored for Retrieval-Augmented Generation (RAG) evaluation. Our benchmark introduces robustness evaluation through \emph{\textbf{multi-level query rephrasing}}, further improving upon previous benchmarks.
The benchmark construction begins with \emph{\textbf{document collection}}, followed by four key steps: (1) \emph{\textbf{Query Generation}}, (2) \emph{\textbf{Query Verification}}, (3) \emph{\textbf{Query Rephrasing}}, and (4) \emph{\textbf{False Negative Verification}}.

\subsection{Document Collection}
To reflect real-world retrieval challenges, we focus on \emph{\textbf{long documents}} rather than isolated pages, and also ensuring \emph{\textbf{many pages within the same sub-domain}} by focusing on a single company data (IBM). Our dataset consists of ~8000 pages across four sub-domains, forming four specialized benchmarks (see \cref{Table:benchmark_statistics} for details). For each page, we added the document name to the page image to provide context.
\textbf{FinReport}: Financial reports (2005--2023), totaling 19 documents and 2687 pages, with a mix of text and tables.  
\textbf{FinSlides}: Quarterly financial presentations (2008--2024), totaling 65 presentations and 2280 pages, primarily table-heavy.  
\textbf{TechReport}: 17 Technical documents on FlashSystem, totaling 1674 pages, text-heavy with visual elements and tables. 
\textbf{TechSlides}: 62 Technical presentations on business and IT automation, totaling 1963 pages, with significant visual content.

\subsection{Query Generation \& Filtering}
\paragraph{Generation.} We aim to generate queries that are both answerable by a specific document and RAG-suitable, meaning they reflect natural user inquiries without prior knowledge of the exact page or answer location (unlike traditional Q/A datasets tied to specific pages). 
To achieve this, we employed a Pixtral-12B VLM~\citep{agrawal2024pixtral}, prompting it to generate RAG-specific questions (see \cref{fig:query_generation_prompt}). Each document page was fed into the VLM, which produced 10 query-answer pairs per page, later keeping only a subset that met the benchmark’s quality criteria after filtering. Each retained query-answer pair is labeled with the corresponding page it was generated from.

\paragraph{Verification.}
Although the VLM is instructed to generate RAG-specific queries, many still do not fully align with our requirements. To systematically classify them, we use Mixtral-8x22B-v0.1 LLM \citep{jiang2024mixtral}, which evaluates each generated query and determines whether it is suitable as retrieval query  (see prompt in \cref{fig:query_verification_prompt}).
Queries that are well-formed for RAG are those that a user might ask without prior knowledge of the document’s structure, ensuring they are neither too general nor overly specific to a single page. Queries that fail this criterion fall into two categories: those with explicit page references, such as "in Figure 5" or "the title of the page", and those that are too broad, like "What is the net revenue in 2020?" instead of "What is IBM’s net revenue in 2020?".

\subsection{Query Rephrasing}
In real-world retrieval, a user formulating a query does not have direct access to the document’s content and will naturally phrase their question without mirroring the exact wording from the source. However, VLMs often generate queries by copying phrases directly, leading to an over-reliance on keyword matching rather than true semantic retrieval. To address this, we introduce a rephrasing step that preserves query meaning while reducing dependence on specific document wording.  Each query is processed by Mixtral-8x22B-v0.1 with a dedicated prompt designed to alter phrasing while maintaining intent.  
The rephrased query is then verified by the LLM using a validation prompt (\cref{fig:rephrasing_prompts}), along with the original query and answer, to ensure it retains the original meaning and still corresponds to the known answer in the labeled page.  

\vspace{0.1cm}
\noindent
To enable deeper evaluation, each query undergoes three levels of rephrasing using distinct prompts (\cref{fig:rephrasing_prompts}). The first level introduces minor word changes while maintaining structure. The second modifies word choice and sentence order, making the phrasing more distinct. The third involves significant word rephrasing and sentence restructuring while preserving meaning. At the end of this process, each query exists in four versions: the original and three progressively rephrased forms, all linked to the same document page (see examples in \cref{fig:ours_examples_1,fig:ours_examples_2}).

\subsection{Accurate Labeling}

The final step in preparing our benchmark is verifying the correctness of negative labels.  This is especially crucial for our challenging benchmarks, where many pages share highly similar content within the same sub-domain.
Each query is systematically tested against all benchmark pages. Though computationally expensive, this step prevents false negatives and ensures reliable evaluation. Queries together with each page are processed using Pixtral-12B, which determines whether a page contains an answer to the query. Every query is then explicitly linked to all relevant pages. For simplicity, our final benchmark retains queries whose only the originally assigned page is verified to contain the correct answer. This results in a high-quality dataset of triplets: a page image, a query, and its corresponding answer. Note that our benchmark includes pages without corresponding queries. These are pages whose queries were filtered out at some stage, either because they were not suitable for RAG-style questions in general (e.g., title pages) or because the specific generated queries were not suitable for RAG.

\section{Benchmarks Quality Evaluation} \label{sec:Benchmarks_Comparisons}

A high-quality benchmark for multi-modal retrieval is essential, yet few existing benchmarks are designed for this purpose, and none comprehensively define or implement the necessary properties. \cref{Table:benchmark_comparison} compares our benchmark with other prominent ones, which suffer from limitations such as poor alignment with real-world queries, high false-negative rates, and trivial difficulty.

\paragraph{Accurate Labeling.}
Many perceived retrieval errors in existing benchmarks are actually false negatives, meaning pages that correctly answer the query but were mislabeled as irrelevant. To mitigate this, we introduce a false-negative verification process that exhaustively labels all valid pages.
\textbf{\emph{Human Evaluation.}} We sampled 50 top-1 retrieval errors of ColQwen on Vidore, MMlongbench, and our benchmark.
Annotators reviewed the query and retrieved page (labeled as negative) to determine if it could answer the query (\cref{fig:Human_False_Negative}). A total of 234 responses from 5 annotators were collected.

\vspace{0.15cm}
\hspace{0.2cm}
    {\footnotesize
\centering
\begin{tabular}{lccc} 
\toprule & \textbf{Vidore} & \textbf{MMLong}  & \textbf{Ours}  \\
\midrule
\textbf{False Negative (\%) \(\downarrow\)} & 86.9 & 77.8 & 31.9 \\
\bottomrule
\end{tabular}}

\vspace{0.15cm}
\noindent
The table shows that Vidore and MMlongbench had a high rate of false negatives, whereas our benchmark, despite its challenging design with similar sub-domain pages, had significantly fewer, proving the effectiveness of accurate labeling.

\paragraph{Enhanced Difficulty.}
A strong benchmark must pose real challenges. Existing ones fall short by offering too few relevant candidates or allowing retrieval via simple keyword matching rather than true semantic understanding.  
For example, having 1,000 financial pages from different companies is insufficient, as knowing the company name narrows the candidates to a few dozen. The ColQwen model achieves an NDCG@5 of around 90 on Vidore. Other sub-datasets, although reporting lower performance, contain many errors that are actually false negatives, as demonstrated by our human evaluation presented above.  
We address this issue through accurate labeling and by incorporating long documents and extensive sub-domain coverage. This provides many similar pages, making retrieval more challenging and better reflecting real-world scenarios. Moreover, we prevent trivial keyword-based retrieval by introducing the first rephrasing benchmark for multi-modal document RAG, ensuring robustness to query variations and promoting semantic learning.

\paragraph{Realistic-RAG Queries.}
To reflect real RAG use cases, queries must resemble natural information-seeking questions. Our benchmark ensures this through a two-step RAG-tailored pipeline: generation and filtering. 
\textbf{\emph{Human Evaluation.}} We randomly sampled 50 queries from Vidore, MMLongBench, and our benchmark.
Annotators, unaware of the source benchmark or study goal, evaluated whether each query could reasonably be asked by a real user (\cref{fig:Human_query_RAG}). A total of 578 responses were collected from 5 annotators.

{\footnotesize
\vspace{0.15cm}
\hspace{0.02cm}
\centering
\begin{tabular}{@{\extracolsep{\fill}}lccc} 
\toprule & \textbf{Vidore} & \textbf{MMLong}  & \textbf{Ours}  \\
\midrule
\textbf{Realistic-RAG Queries (\%) \(\uparrow\) } & 43.6  & 35.2  & 85.0 \\
\bottomrule
\end{tabular}}

\vspace{0.15cm}
\noindent
The table shows that most Vidore/MMLongBench queries were labeled as unrealistic RAG queries (see some examples in \cref{fig:others_examples}), whereas 85\% of ours were validated as realistic, highlighting shortcomings in existing benchmarks and the effectiveness of our query generation and filtering process.

\begin{table}[ht!]
\footnotesize
\renewcommand{\arraystretch}{1.5} 
\setlength{\dashlinedash}{0.5pt}
\setlength{\dashlinegap}{0.5pt}
\setlength\tabcolsep{4pt} 

\hspace{-0.15cm}
\begin{tabular}{@{\extracolsep{\fill}}lcccc} 
\toprule
\textbf{Benchmark} & \textbf{FinReport} & \textbf{FinSlides} & \textbf{TechReport} & \textbf{TechSlides}  \\
\midrule
\textbf{\emph{\underline{Text}}} \\
\textit{BM25 (OCR)}    & 21.7 & 5.9 & 35.1 & 31.2   \\
\textit{BGE-M3 (OCR)}  & 36.5 & 11.4 & 37.1 & 49.7  \\
\textit{BM25 (Captioning)}  & 25.3 & 9.9 & 37.2 & 36.1    \\
\textit{BGE-M3 (Captioning)} & 35.9 & 13.8 & 37.5 & 51.7   \\
\midrule
 \underline{\emph{\textbf{Vision}}} \\
\textit{ColPali}  & 34.5 & 27.6 & 62.0 & 75.8  \\
\textit{\textbf{Rob}ColPali}  & 47.1 {\tiny \textcolor{DarkGreen}{+12.6}}  & 48.4 {\tiny \textcolor{DarkGreen}{+20.8}}  & 66.6 {\tiny \textcolor{DarkGreen}{+4.60}}  & 82.8 {\tiny \textcolor{DarkGreen}{+7.0}}  \\
\textit{\textbf{Tab}ColPali}  & 50.5 {\tiny \textcolor{DarkGreen}{+16.0}}  & 41.5 {\tiny \textcolor{DarkGreen}{+13.9}}  & 61.3 {\tiny \textcolor{red}{-0.7}}  & 77.6 {\tiny \textcolor{DarkGreen}{+1.8}}  \\
\textit{\textbf{RobTab}ColPali}  & \textbf{63.2} {\tiny \textcolor{DarkGreen}{+28.7}}  & \textbf{58.3} {\tiny \textcolor{DarkGreen}{+30.7}}  & \textbf{70.7} {\tiny \textcolor{DarkGreen}{+8.7}}  & \textbf{83.3} {\tiny \textcolor{DarkGreen}{+7.5}}  \\
\addlinespace
\textit{ColQwen}   & 41.8 & 31.1 & 66.9 & 78.1  \\
\textit{\textbf{Rob}ColQwen}  & 47.5 {\tiny \textcolor{DarkGreen}{+5.7}}  & 44.3 {\tiny \textcolor{DarkGreen}{+13.2}}  & 69.5 {\tiny \textcolor{DarkGreen}{+2.6}}  & 83.0 {\tiny \textcolor{DarkGreen}{+4.9}}  \\
\textit{\textbf{Tab}ColQwen}  & 54.0 {\tiny \textcolor{DarkGreen}{+12.2}}  & 49.6 {\tiny \textcolor{DarkGreen}{+18.5}}  & 65.9 {\tiny \textcolor{red}{-1.0}}  & 78.9 {\tiny \textcolor{red}{-0.8}}  \\
\textit{\textbf{RobTab}ColQwen}  & \textbf{\underline{67.1}} {\tiny \textcolor{DarkGreen}{+25.3}}  & \textbf{\underline{61.6}} {\tiny \textcolor{DarkGreen}{+30.5}}  & \textbf{\underline{73.2}} {\tiny \textcolor{DarkGreen}{+6.3}}  & \textbf{\underline{85.0}} {\tiny \textcolor{DarkGreen}{+6.9}}  \\

\bottomrule
\end{tabular}
\caption{
\textbf{Performance of Different Models on Our Benchmark.}  
We evaluate various models, including text- and vision-based approaches, across our four benchmarks. Results, measured using NDCG@5, are reported on our final benchmark with queries rephrased at the highest level (Level 3). We also present results for our fine-tuned models trained on our proposed datasets: \emph{Rob} – trained on a rephrased dataset, \emph{Tab} – trained on a table-heavy dataset, and \emph{RobTab} – incorporating both.
}
\label{Table:model_comparisons_on_benchmarks}
\vspace{-0.4cm}
\end{table}

\paragraph{Summary.}
Our benchmark enhances multi-modal retrieval evaluation by introducing non-trivial difficulty with long documents and broad sub-domain coverage. It ensures RAG-aligned queries and promotes semantic retrieval over keyword matching through query rephrasing, addressing key limitations of existing benchmarks.

\section{Model Evaluation \& Enhancement}
\label{sec:results}

\paragraph{Evaluation Models and Metrics.}
We evaluate multiple models on our benchmark, covering both text and vision-based approaches. We use the ColPali benchmark code (ViDoRe) to assess our text-based models and the vision-based models ColPali and ColQwen.
For the text-based methods, we follow the framework suggested in ColPali, which employs
\emph{Unstructured}
in a high-resolution configuration with OCR engine to parse PDFs.
For each document, \emph{Unstructured} produces text chunks and visual chunks (e.g., tables, figures, images).
We consider two text-based variants:
(1) \emph{OCR}, where visual data is processed through an OCR engine,
and (2) \emph{Captioning}, where visual elements are described using a Vision Language Model (Qwen2-VL-72B-Instruct \cite{wang2024qwen2}).
We then evaluate two retrieval methods:
\emph{Okapi BM25} and \emph{BGE-M3}~\cite{chen2024bge}
 (see \cref{sec:appendix_model_evalaution} for more details).
We report NDCG@5 as our primary ranking metric, which evaluates how well relevant items are ranked within the top 5 results, giving higher importance to those appearing earlier. Additional metrics and details provided in \ref{sec:appendix_model_evalaution} \& \ref{sec:additional_results}.

\subsection{Results Analysis}
In \cref{Table:model_comparisons_on_benchmarks}, we report NDCG@5 performance for different models across our four benchmarks with high rephrasing levels, which better reflect real-world scenarios. We first present observations about the vanilla models, including text-based models, ColPali, and ColQwen:
\emph{\textbf{Visual vs. Text-Based Models.} }
Vision-based models, which use VLMs on page images, significantly outperform text-based models across all benchmarks. This supports the notion that visual information is essential for our benchmark and that these models can effectively utilize it.  
\emph{\textbf{Non-Trivial Difficulty.}} Performance is generally low, especially compared to Vidore, where ColQwen achieves nearly 90\% on average.
\emph{\textbf{Rephrasing Effects on Performance.}} 
Some of the drop in performance is due to rephrasing.
In \cref{Table:average_rephrasing_benchmarks}, we analyze the impact of rephrasing level, showing a clear performance drop as rephrasing intensity increases. BM25 suffers the most, as expected for a lexical-based model, while dense retrieval models are more resilient.
\emph{\textbf{Table-Heavy Finance Benchmarks Are Harder.}}  
Our financial benchmarks (FinReport, FinSlides), which are table-heavy, are significantly more challenging than text/visual-based ones (see table vs. non-table analysis in \cref{Table:model_comparisons_on_labels}).

\subsection{Table and Finance Focused Training}  
To address the challenges in table-heavy datasets, we curated a table-focused finance dataset using FinTabNet~\citep{zheng2020global}, which contains complex tables from S\&P 500 company reports. Through the pipeline in \cref{sec:benchmark}, we generated 46,000 query-answer-page triplets to enhance retrieval for table-heavy financial data (see \cref{Table:VLM_ablation} for a VLM ablation study). We fine-tuned ColPali and ColQwen for one epoch on this dataset while incorporating the ColPali training set, producing the \textbf{\emph{TabCol}} models.

\begin{figure}[h]
    \centering
    \includegraphics[width=0.48\textwidth]{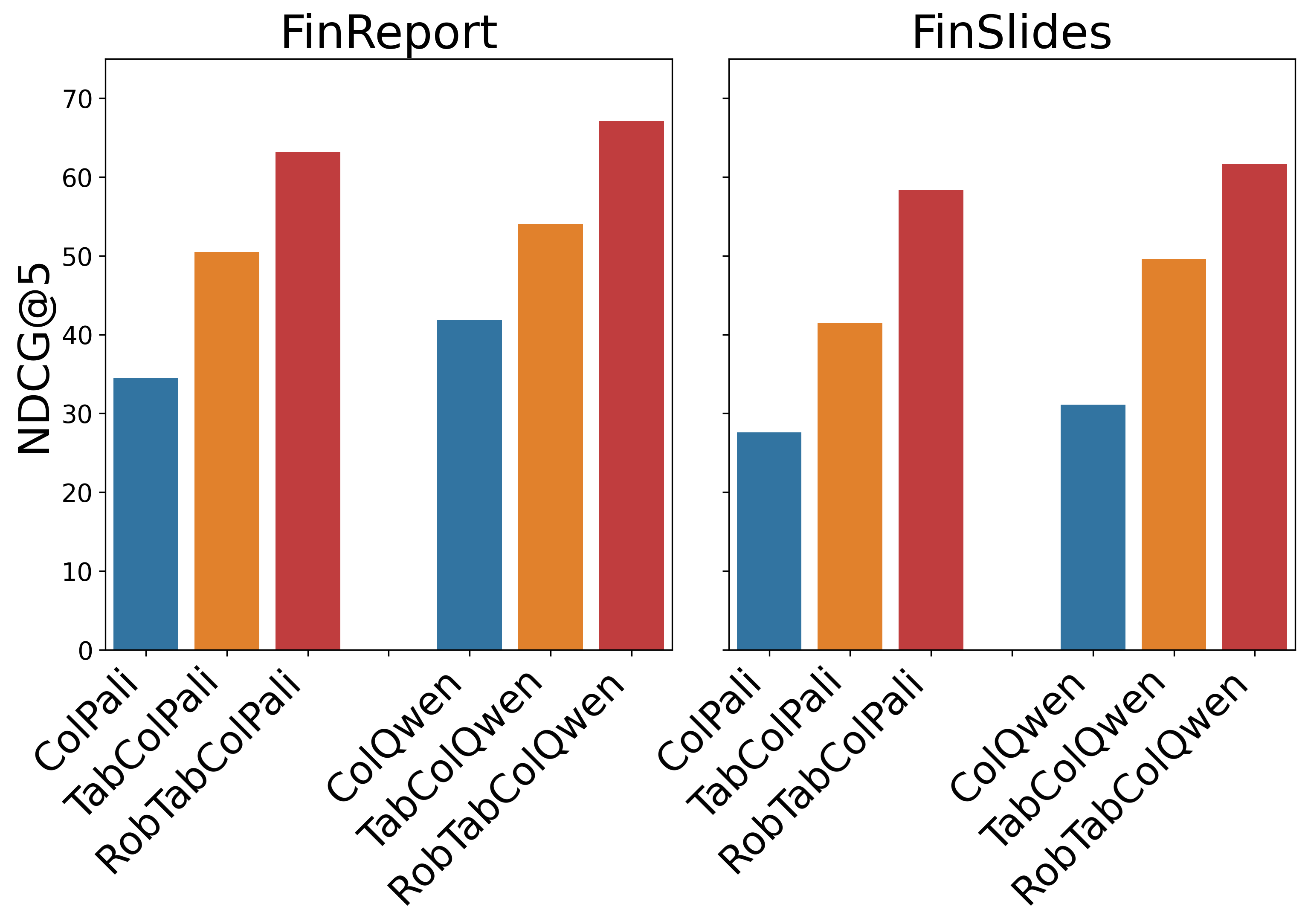} 
    \caption{\textbf{Table-Focused Training Improves Financial Benchmarks.} Fine-tuning with our proposed table-heavy training set, combined with the ColPali training set (both in their original and rephrased versions) significantly enhances performance on financial benchmarks (results shown for rephrasing level 3).}
    \vspace{-0.1cm}
    \label{fig:Table_results_plots} 
\end{figure}

\begin{figure*}[ht]
    \centering
    \begin{minipage}{0.4\textwidth}
        \footnotesize
        \renewcommand{\arraystretch}{1.5} 
        \setlength\tabcolsep{5pt} 

        \vspace{-0.15cm}
        \centering
        \begin{tabular}{lcccc}
            \toprule
            \textbf{Rephrasing Levels} & \textbf{0} & \textbf{1} & \textbf{2} & \textbf{3} \\
            \midrule
            BM25 (Captioning) & 52.7 & 41.6 & 31.3 & 27.1 \\
            BGE-M3 (Captioning) & 43.3 & 39.2 & 36.5 & 35.2 \\
            \addlinespace
            ColPali & 71.3 & 65.3 & 60.3 & 56.6 \\
            \textbf{Rob}ColPali & 76.7 & 72.6 & 68.4 & 65.7 \\
            \textbf{RobTab}ColPali & 80.8 & 77.8 & 74.9 & 72.7 \\
            \addlinespace
            ColQwen & 78.9 & 72.5 & 68.2 & 65.3 \\
            \textbf{Rob}ColQwen & 81.6 & 77.3 & 73.7 & 71.0 \\
            \textbf{RobTab}ColQwen & 85.1 & 81.7 & 78.6 & 76.4 \\
            \bottomrule
        \end{tabular}
        \vspace{-0.05cm}
        \captionof{table}{
        \textbf{Query Rephrasing Effect.} This table presents the NDCG@5 scores averaged across all benchmarks for different models and rephrasing levels. `0' represents no rephrasing, while `3' indicates significant rephrasing (see \cref{Table:rephrasing_benchmarks} for full results).}
        \label{Table:average_rephrasing_benchmarks}
    \end{minipage}
    \hfill
    \begin{minipage}{0.58\textwidth}
        \centering
        \includegraphics[width=\linewidth]{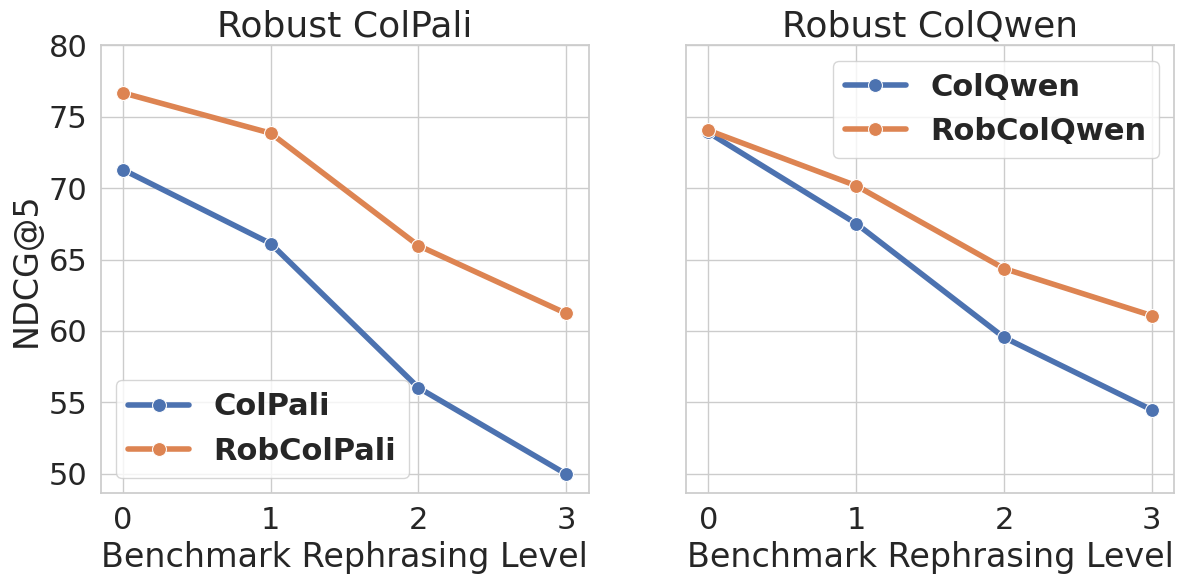}
        
        \begin{minipage}{0.95\linewidth} 
            \centering
            \vspace{0.41cm}
            \caption{\textbf{Fine-Tuning on Rephrased Training Set.} We compare the NDCG@5 scores across rephrasing levels for baseline models (ColPali and ColQwen) against our fine-tuned models (RobCol). The results demonstrate that fine-tuning with our rephrased training data significantly enhances rephrasing robustness for both ColPali and ColQwen.}
            \label{fig:rephrasing_table_plot}
        \end{minipage}
    \end{minipage}
    \vspace{-0.4cm}
\end{figure*}

\vspace{0.1cm}
\noindent
As shown in \cref{fig:Table_results_plots} and \cref{Table:model_comparisons_on_benchmarks}, TabCol models significantly improve performance on financial benchmarks, effectively addressing table-heavy dataset challenges. Importantly, this enhancement does not come at the cost of generalization, as TabCol models continue well across the two other benchmarks. 

\vspace{0.1cm}

\subsection{Rephrasing Robustness Training}  
Our benchmark reveals that current models struggle with rephrasing, suggesting that training and evaluation queries often closely match the phrasing of their retrieved pages. To address this, we augmented the ColPali training set by rephrasing half of its queries, randomly selecting one of three rephrasing levels. This was done using 
LLaMA-3-70B\footnote{\url{https://huggingface.co/meta-llama/Llama-3-70B-Instruct}}
, a different LLM than the one used for the benchmark. 
This dataset forces models to learn semantics rather than relying on keyword matching. We fine-tuned ColPali-v1.2 and 
ColQwen2-v1.0\footnote{\url{https://huggingface.co/vidore/colqwen2-v1.0}}
(using the ColPali code) for one epoch, producing the RobCol models.

\noindent
As demonstrated in \cref{fig:rephrasing_table_plot,Table:average_rephrasing_benchmarks} (and in \cref{Table:model_comparisons_on_benchmarks,Table:rephrasing_benchmarks}),
RobCol significantly outperforms baselines on rephrased queries while maintaining comparable performance on non-rephrased cases, achieving an average NDCG@5 improvement of 11.1 at rephrasing level 3. 
The gains are stronger on financial benchmarks, where the task is more complex, suggesting that enhanced semantic understanding through rephrasing robustness is particularly beneficial. In \cref{Table:Training_Set_ablation}, we show that fine-tuning on the original data (rather than the rephrased version) provides much lower improvement for ColPali and leads to a decrease in performance for ColQwen.

\noindent
Additionally, we created a rephrased version of the tabled-focused dataset and fine-tuned both models on it, along with the rephrased ColPali training set, producing RobTabCol models. RobTabCol consistently outperforms all models across nearly all benchmarks and rephrasing levels, achieving a 25–30 NDCG@5 improvement over base models on rephrased finance benchmarks and 6–9 on non-financial ones.

\section{Conclusions}
\label{sec:conclusions}

We introduced \emph{REAL-MM-RAG-Bench}, a real-world multi-modal retrieval benchmark designed to evaluate retrieval models in \emph{reliable, challenging, and realistic} settings. Our benchmark addresses key properties essential for evaluating retrieval systems in real-world applications, which prior benchmarks often fail to capture. An important contribution of our work is the introduction of a \emph{multi-level rephrasing evaluation}, which assesses models under increasing linguistic variation, highlighting their limitations in generalizing beyond surface-level text matching.

\vspace{0.1cm}
\noindent
Our findings reveal two major weaknesses in current models: (i) retrieval over table-heavy financial documents and (ii) sensitivity to query rephrasing. To address these, we proposed dedicated training sets: a \emph{finance-table-heavy dataset} to improve retrieval on tabular content and a \emph{rephrased dataset} to enhance model robustness to query variations. Fine-tuning on these datasets yields significant improvements across benchmarks, demonstrating the impact of targeted training data. \emph{REAL-MM-RAG-Bench} and our proposed solutions establish a foundation for future research, paving the way for robust and effective retrieval models in real-world multi-modal retrieval scenarios.

\clearpage
\section{Limitations} \label{sec:limitations}

While REAL-MM-RAG-Bench presents a comprehensive and realistic evaluation framework for multi-modal retrieval, several limitations remain:

\noindent
\emph{Query Variability:} Our queries are generated using a Vision-Language Model (VLM), which, while effective, may not fully capture the full range of plausible user queries.

\noindent
\emph{LLM and VLM Limitations:} Despite the strength of modern Large Language Models (LLMs) and VLMs, our filtering strategies and labeling process remain subject to their limitations. While our human evaluation confirms the effectiveness of our approach, errors in labeling and query selection may still occur. As LLMs and VLMs continue to improve, future benchmarks could leverage more accurate models to refine dataset construction further.

\noindent
\emph{Multi-Page Reasoning Queries:} Our benchmark is designed to best evaluate the retrieval component of Retrieval-Augmented Generation (RAG). While the dataset can be used for the generation step as well, it does not explicitly assess multi-page reasoning. Future work could explore automated query generation that combines multiple pages using LLMs and/or VLMs to construct multi-page reasoning tasks, enhancing the benchmark’s ability to evaluate complex retrieval scenarios.

\noindent
REAL-MM-RAG-Bench provides a realistic, reliable, and challenging retrieval benchmark, helping to identify critical weaknesses in current multi-modal retrieval models and paving the way for future improvements in both evaluation and model development.

\bibliography{custom}

\begin{thebibliography}{40}
\providecommand{\natexlab}[1]{#1}

\bibitem[{Agrawal et~al.(2024)Agrawal, Antoniak, Hanna, Bout, Chaplot, Chudnovsky, Costa, De~Monicault, Garg, Gervet et~al.}]{agrawal2024pixtral}
Pravesh Agrawal, Szymon Antoniak, Emma~Bou Hanna, Baptiste Bout, Devendra Chaplot, Jessica Chudnovsky, Diogo Costa, Baudouin De~Monicault, Saurabh Garg, Theophile Gervet, et~al. 2024.
\newblock Pixtral 12b.
\newblock \emph{arXiv preprint arXiv:2410.07073}.

\bibitem[{Bailey et~al.(2016)Bailey, Moffat, Scholer, and Thomas}]{bailey2016uqv100}
Peter Bailey, Alistair Moffat, Falk Scholer, and Paul Thomas. 2016.
\newblock Uqv100: A test collection with query variability.
\newblock In \emph{Proceedings of the 39th International ACM SIGIR conference on Research and Development in Information Retrieval}, pages 725--728.

\bibitem[{Bailey et~al.(2017)Bailey, Moffat, Scholer, and Thomas}]{bailey2017retrieval}
Peter Bailey, Alistair Moffat, Falk Scholer, and Paul Thomas. 2017.
\newblock Retrieval consistency in the presence of query variations.
\newblock In \emph{Proceedings of the 40th international ACM SIGIR conference on research and development in information retrieval}, pages 395--404.

\bibitem[{Benham et~al.(2018)Benham, Culpepper, Gallagher, Lu, and Mackenzie}]{benham2018towards}
Rodger Benham, J~Shane Culpepper, Luke Gallagher, Xiaolu Lu, and Joel~M Mackenzie. 2018.
\newblock Towards efficient and effective query variant generation.
\newblock In \emph{DESIRES}, pages 62--67.

\bibitem[{Beyer et~al.(2024)Beyer, Steiner, Pinto, Kolesnikov, Wang, Salz, Neumann, Alabdulmohsin, Tschannen, Bugliarello et~al.}]{beyer2024paligemma}
Lucas Beyer, Andreas Steiner, Andr{\'e}~Susano Pinto, Alexander Kolesnikov, Xiao Wang, Daniel Salz, Maxim Neumann, Ibrahim Alabdulmohsin, Michael Tschannen, Emanuele Bugliarello, et~al. 2024.
\newblock Paligemma: A versatile 3b vlm for transfer.
\newblock \emph{arXiv preprint arXiv:2407.07726}.

\bibitem[{Chen et~al.(2024)Chen, Xiao, Zhang, Luo, Lian, and Liu}]{chen2024bge}
Jianlv Chen, Shitao Xiao, Peitian Zhang, Kun Luo, Defu Lian, and Zheng Liu. 2024.
\newblock Bge m3-embedding: Multi-lingual, multi-functionality, multi-granularity text embeddings through self-knowledge distillation.
\newblock \emph{arXiv preprint arXiv:2402.03216}.

\bibitem[{Devlin(2018)}]{devlin2018bert}
Jacob Devlin. 2018.
\newblock Bert: Pre-training of deep bidirectional transformers for language understanding.
\newblock \emph{arXiv preprint arXiv:1810.04805}.

\bibitem[{Ding et~al.(2024)Ding, Ren, Huang, Luo, and Han}]{ding2024mvqa}
Yihao Ding, Kaixuan Ren, Jiabin Huang, Siwen Luo, and Soyeon~Caren Han. 2024.
\newblock Mvqa: A dataset for multimodal information retrieval in pdf-based visual question answering.
\newblock \emph{arXiv preprint arXiv:2404.12720}.

\bibitem[{Faysse et~al.(2024)Faysse, Sibille, Wu, Omrani, Viaud, Hudelot, and Colombo}]{faysse2024colpali}
Manuel Faysse, Hugues Sibille, Tony Wu, Bilel Omrani, Gautier Viaud, C{\'e}line Hudelot, and Pierre Colombo. 2024.
\newblock Colpali: Efficient document retrieval with vision language models.
\newblock \emph{arXiv preprint arXiv:2407.01449}.

\bibitem[{Formal et~al.(2021)Formal, Piwowarski, and Clinchant}]{formal2021splade}
Thibault Formal, Benjamin Piwowarski, and St{\'e}phane Clinchant. 2021.
\newblock Splade: Sparse lexical and expansion model for first stage ranking.
\newblock In \emph{Proceedings of the 44th International ACM SIGIR Conference on Research and Development in Information Retrieval}, pages 2288--2292.

\bibitem[{Hagen et~al.(2024)Hagen, Scells, and Potthast}]{hagen2024revisiting}
Tim Hagen, Harrisen Scells, and Martin Potthast. 2024.
\newblock Revisiting query variation robustness of transformer models.
\newblock In \emph{Findings of the Association for Computational Linguistics: EMNLP 2024}, pages 4283--4296.

\bibitem[{Islam et~al.(2023)Islam, Kannappan, Kiela, Qian, Scherrer, and Vidgen}]{islam2023financebench}
Pranab Islam, Anand Kannappan, Douwe Kiela, Rebecca Qian, Nino Scherrer, and Bertie Vidgen. 2023.
\newblock Financebench: A new benchmark for financial question answering.
\newblock \emph{arXiv preprint arXiv:2311.11944}.

\bibitem[{Jiang et~al.(2024)Jiang, Sablayrolles, Roux, Mensch, Savary, Bamford, Chaplot, Casas, Hanna, Bressand et~al.}]{jiang2024mixtral}
Albert~Q Jiang, Alexandre Sablayrolles, Antoine Roux, Arthur Mensch, Blanche Savary, Chris Bamford, Devendra~Singh Chaplot, Diego de~las Casas, Emma~Bou Hanna, Florian Bressand, et~al. 2024.
\newblock Mixtral of experts.
\newblock \emph{arXiv preprint arXiv:2401.04088}.

\bibitem[{Karpukhin et~al.(2020)Karpukhin, O{\u{g}}uz, Min, Lewis, Wu, Edunov, Chen, and Yih}]{karpukhin2020dense}
Vladimir Karpukhin, Barlas O{\u{g}}uz, Sewon Min, Patrick Lewis, Ledell Wu, Sergey Edunov, Danqi Chen, and Wen-tau Yih. 2020.
\newblock Dense passage retrieval for open-domain question answering.
\newblock \emph{arXiv preprint arXiv:2004.04906}.

\bibitem[{Khattab and Zaharia(2020)}]{khattab2020colbert}
Omar Khattab and Matei Zaharia. 2020.
\newblock Colbert: Efficient and effective passage search via contextualized late interaction over bert.
\newblock In \emph{Proceedings of the 43rd International ACM SIGIR conference on research and development in Information Retrieval}, pages 39--48.

\bibitem[{Lewis et~al.(2020)Lewis, Perez, Piktus, Petroni, Karpukhin, Goyal, K{\"u}ttler, Lewis, Yih, Rockt{\"a}schel et~al.}]{lewis2020retrieval}
Patrick Lewis, Ethan Perez, Aleksandra Piktus, Fabio Petroni, Vladimir Karpukhin, Naman Goyal, Heinrich K{\"u}ttler, Mike Lewis, Wen-tau Yih, Tim Rockt{\"a}schel, et~al. 2020.
\newblock Retrieval-augmented generation for knowledge-intensive nlp tasks.
\newblock \emph{Advances in Neural Information Processing Systems}, 33:9459--9474.

\bibitem[{Lu et~al.(2019)Lu, Kurland, Culpepper, Craswell, and Rom}]{lu2019relevance}
Xiaolu Lu, Oren Kurland, J~Shane Culpepper, Nick Craswell, and Ofri Rom. 2019.
\newblock Relevance modeling with multiple query variations.
\newblock In \emph{Proceedings of the 2019 ACM SIGIR International Conference on Theory of Information Retrieval}, pages 27--34.

\bibitem[{Ma et~al.(2024{\natexlab{a}})Ma, Lin, Li, Chen, and Lin}]{ma2024unifying}
Xueguang Ma, Sheng-Chieh Lin, Minghan Li, Wenhu Chen, and Jimmy Lin. 2024{\natexlab{a}}.
\newblock Unifying multimodal retrieval via document screenshot embedding.
\newblock \emph{arXiv preprint arXiv:2406.11251}.

\bibitem[{Ma et~al.(2024{\natexlab{b}})Ma, Zang, Chen, Chen, Jiao, Li, Lu, Liu, Ma, Dong et~al.}]{ma2024mmlongbench}
Yubo Ma, Yuhang Zang, Liangyu Chen, Meiqi Chen, Yizhu Jiao, Xinze Li, Xinyuan Lu, Ziyu Liu, Yan Ma, Xiaoyi Dong, et~al. 2024{\natexlab{b}}.
\newblock Mmlongbench-doc: Benchmarking long-context document understanding with visualizations.
\newblock \emph{arXiv preprint arXiv:2407.01523}.

\bibitem[{Masry et~al.(2022)Masry, Long, Tan, Joty, and Hoque}]{masry2022chartqa}
Ahmed Masry, Do~Xuan Long, Jia~Qing Tan, Shafiq Joty, and Enamul Hoque. 2022.
\newblock Chartqa: A benchmark for question answering about charts with visual and logical reasoning.
\newblock \emph{arXiv preprint arXiv:2203.10244}.

\bibitem[{Mathew et~al.(2021)Mathew, Karatzas, and Jawahar}]{mathew2021docvqa}
Minesh Mathew, Dimosthenis Karatzas, and CV~Jawahar. 2021.
\newblock Docvqa: A dataset for vqa on document images.
\newblock In \emph{Proceedings of the IEEE/CVF winter conference on applications of computer vision}, pages 2200--2209.

\bibitem[{Penha et~al.(2022)Penha, C{\^a}mara, and Hauff}]{penha2022evaluating}
Gustavo Penha, Arthur C{\^a}mara, and Claudia Hauff. 2022.
\newblock Evaluating the robustness of retrieval pipelines with query variation generators.
\newblock In \emph{European conference on information retrieval}, pages 397--412. Springer.

\bibitem[{Radford et~al.(2021)Radford, Kim, Hallacy, Ramesh, Goh, Agarwal, Sastry, Askell, Mishkin, Clark et~al.}]{radford2021learning}
Alec Radford, Jong~Wook Kim, Chris Hallacy, Aditya Ramesh, Gabriel Goh, Sandhini Agarwal, Girish Sastry, Amanda Askell, Pamela Mishkin, Jack Clark, et~al. 2021.
\newblock Learning transferable visual models from natural language supervision.
\newblock In \emph{International conference on machine learning}, pages 8748--8763. PMLR.

\bibitem[{Raffel et~al.(2020)Raffel, Shazeer, Roberts, Lee, Narang, Matena, Zhou, Li, and Liu}]{raffel2020exploring}
Colin Raffel, Noam Shazeer, Adam Roberts, Katherine Lee, Sharan Narang, Michael Matena, Yanqi Zhou, Wei Li, and Peter~J Liu. 2020.
\newblock Exploring the limits of transfer learning with a unified text-to-text transformer.
\newblock \emph{Journal of machine learning research}, 21(140):1--67.

\bibitem[{Ramos et~al.(2023)Ramos, Elliott, and Martins}]{ramos2023retrieval}
Rita Ramos, Desmond Elliott, and Bruno Martins. 2023.
\newblock Retrieval-augmented image captioning.
\newblock \emph{arXiv preprint arXiv:2302.08268}.

\bibitem[{Robertson et~al.(1994)Robertson, Walker, Jones, and GATFORD}]{robertson1994okapi}
S~Robertson, Steve Walker, Susan Jones, and MHB GATFORD. 1994.
\newblock Okapi at 3.
\newblock In \emph{Proceedings of the 3rd Text REtrieval Conference (-3)}, pages 109--126.

\bibitem[{Sidiropoulos and Kanoulas(2022)}]{sidiropoulos2022analysing}
Georgios Sidiropoulos and Evangelos Kanoulas. 2022.
\newblock Analysing the robustness of dual encoders for dense retrieval against misspellings.
\newblock In \emph{Proceedings of the 45th International ACM SIGIR Conference on Research and Development in Information Retrieval}, pages 2132--2136.

\bibitem[{Smeaton and Kelledy(1998)}]{smeaton1998user}
Alan~F Smeaton and Fergus Kelledy. 1998.
\newblock User-chosen phrases in interactive query formulation for information retrieval.
\newblock In \emph{20th Annual BCS-IRSG Colloquium on IR}. BCS Learning \& Development.

\bibitem[{Smith(2007)}]{smith2007overview}
Ray Smith. 2007.
\newblock An overview of the tesseract ocr engine.
\newblock In \emph{Ninth international conference on document analysis and recognition (ICDAR 2007)}, volume~2, pages 629--633. IEEE.

\bibitem[{Sparck~Jones(1972)}]{sparck1972statistical}
Karen Sparck~Jones. 1972.
\newblock A statistical interpretation of term specificity and its application in retrieval.
\newblock \emph{Journal of documentation}, 28(1):11--21.

\bibitem[{Tanaka et~al.(2023)Tanaka, Nishida, Nishida, Hasegawa, Saito, and Saito}]{tanaka2023slidevqa}
Ryota Tanaka, Kyosuke Nishida, Kosuke Nishida, Taku Hasegawa, Itsumi Saito, and Kuniko Saito. 2023.
\newblock Slidevqa: A dataset for document visual question answering on multiple images.
\newblock In \emph{Proceedings of the AAAI Conference on Artificial Intelligence}, volume~37, pages 13636--13645.

\bibitem[{Wang et~al.(2024{\natexlab{a}})Wang, Bai, Tan, Wang, Fan, Bai, Chen, Liu, Wang, Ge, Fan, Dang, Du, Ren, Men, Liu, Zhou, Zhou, and Lin}]{Qwen2VL}
Peng Wang, Shuai Bai, Sinan Tan, Shijie Wang, Zhihao Fan, Jinze Bai, Keqin Chen, Xuejing Liu, Jialin Wang, Wenbin Ge, Yang Fan, Kai Dang, Mengfei Du, Xuancheng Ren, Rui Men, Dayiheng Liu, Chang Zhou, Jingren Zhou, and Junyang Lin. 2024{\natexlab{a}}.
\newblock Qwen2-vl: Enhancing vision-language model's perception of the world at any resolution.
\newblock \emph{arXiv preprint arXiv:2409.12191}.

\bibitem[{Wang et~al.(2024{\natexlab{b}})Wang, Bai, Tan, Wang, Fan, Bai, Chen, Liu, Wang, Ge et~al.}]{wang2024qwen2}
Peng Wang, Shuai Bai, Sinan Tan, Shijie Wang, Zhihao Fan, Jinze Bai, Keqin Chen, Xuejing Liu, Jialin Wang, Wenbin Ge, et~al. 2024{\natexlab{b}}.
\newblock Qwen2-vl: Enhancing vision-language model's perception of the world at any resolution.
\newblock \emph{arXiv preprint arXiv:2409.12191}.

\bibitem[{Xiong et~al.(2020)Xiong, Xiong, Li, Tang, Liu, Bennett, Ahmed, and Overwijk}]{xiong2020approximate}
Lee Xiong, Chenyan Xiong, Ye~Li, Kwok-Fung Tang, Jialin Liu, Paul Bennett, Junaid Ahmed, and Arnold Overwijk. 2020.
\newblock Approximate nearest neighbor negative contrastive learning for dense text retrieval.
\newblock \emph{arXiv preprint arXiv:2007.00808}.

\bibitem[{Yu et~al.(2024)Yu, Tang, Xu, Cui, Ran, Yan, Liu, Wang, Han, Liu et~al.}]{yu2024visrag}
Shi Yu, Chaoyue Tang, Bokai Xu, Junbo Cui, Junhao Ran, Yukun Yan, Zhenghao Liu, Shuo Wang, Xu~Han, Zhiyuan Liu, et~al. 2024.
\newblock Visrag: Vision-based retrieval-augmented generation on multi-modality documents.
\newblock \emph{arXiv preprint arXiv:2410.10594}.

\bibitem[{Zhai et~al.(2023)Zhai, Mustafa, Kolesnikov, and Beyer}]{zhai2023sigmoid}
Xiaohua Zhai, Basil Mustafa, Alexander Kolesnikov, and Lucas Beyer. 2023.
\newblock Sigmoid loss for language image pre-training.
\newblock In \emph{Proceedings of the IEEE/CVF International Conference on Computer Vision}, pages 11975--11986.

\bibitem[{Zheng et~al.(2021)Zheng, Burdick, Popa, Zhong, and Wang}]{zheng2020global}
Xinyi Zheng, Doug Burdick, Lucian Popa, Peter Zhong, and Nancy Xin~Ru Wang. 2021.
\newblock Global table extractor (gte): A framework for joint table identification and cell structure recognition using visual context.
\newblock \emph{Winter Conference for Applications in Computer Vision (WACV)}.

\bibitem[{Zhu et~al.(2022)Zhu, Lei, Feng, Wang, Zhang, and Chua}]{zhu2022towards}
Fengbin Zhu, Wenqiang Lei, Fuli Feng, Chao Wang, Haozhou Zhang, and Tat-Seng Chua. 2022.
\newblock Towards complex document understanding by discrete reasoning.
\newblock In \emph{Proceedings of the 30th ACM International Conference on Multimedia}, pages 4857--4866.

\bibitem[{Zhu et~al.(2024)Zhu, Huang, Rudinac, and Kanoulas}]{zhu2024enhancing}
Hongyi Zhu, Jia-Hong Huang, Stevan Rudinac, and Evangelos Kanoulas. 2024.
\newblock Enhancing interactive image retrieval with query rewriting using large language models and vision language models.
\newblock In \emph{Proceedings of the 2024 International Conference on Multimedia Retrieval}, pages 978--987.

\bibitem[{Zuccon et~al.(2016)Zuccon, Palotti, and Hanbury}]{zuccon2016query}
Guido Zuccon, Joao Palotti, and Allan Hanbury. 2016.
\newblock Query variations and their effect on comparing information retrieval systems.
\newblock In \emph{Proceedings of the 25th ACM international on conference on information and knowledge management}, pages 691--700.

\end{thebibliography}

\appendix

\clearpage
\section{Appendix}
\label{sec:appendix}

\renewcommand{\thefigure}{S\arabic{figure}} 
\renewcommand{\thetable}{S\arabic{table}}   
\setcounter{figure}{0}  
\setcounter{table}{0}

\subsection{Models Evaluation} \label{sec:appendix_model_evalaution}
Beside the models from ColPali, we have evaluated on text based methods.
Following the ColPali framework, we adopt \emph{Unstructured} as our PDF parser in its high-resolution configuration, which relies on the Tesseract \cite{smith2007overview} OCR engine.
Unstructured processes each document into two main types of chunks: text chunks and visual chunks (e.g., tables, figures, images).
We then construct two text-based variants of our benchmark, differing in how visual chunks are converted into text:
\begin{enumerate}[leftmargin=*]
    \item \textbf{OCR}: Text is retained; tables, figures, and images undergo OCR extraction.
    \item \textbf{Captioning}: Text remains unchanged; visual elements are described using Qwen2.5-VL-72B-Instruct.
\end{enumerate}

\paragraph{Retrieval Methods.} 

We evaluate two retrieval approaches:  
\begin{itemize}[leftmargin=*]
    \item \textbf{Okapi BM25}: A sparse statistical baseline.
    \item \textbf{BGE-M3 (multi-vector)}: A state-of-the-art embedding model.
\end{itemize}

In line with ColPali, chunks are embedded and scored independently, and page-level scores are then obtained via maximum pooling across all chunks for a given page.

\subsection{Proposed Training sets and Fine-tuning} \label{sec:training_Sets_and_finetune_detailes}

\paragraph{Rephrasing Augmentation Training.}
We aimed to fine-tune a trained model with a short training phase to improve robustness to rephrasing. To achieve this, we created a rephrased training set based on the ColPali training data. Specifically, we used approximately half of the full training set (56k queries) and generated rephrased versions. The rephrasing was performed using LLaMA-3-70B, a different LLM than the one used for filtering and rephrasing in the benchmark.

Each query was randomly rephrased using one of three rephrasing levels (different prompts, see \cref{fig:rephrasing_prompts}). To ensure semantic consistency, we used a secondary LLM verification step: the original query and its corresponding answer were fed alongside the rephrased query, and if the meaning was not preserved, the original query was retained in the rephrased training set.

For fine-tuning, we used the full ColPali training set, incorporating the rephrased queries in approximately half of it. The model was trained for one epoch with configurations similar to those used in the original ColPali/ColQwen training.

\paragraph{Table and Finance-Focused Training.}
As we observed that current models performed significantly worse on our table-heavy finance benchmark, we curated a dedicated training set to address this gap. We leveraged the publicly available FinTabNet dataset, a large-scale resource designed for financial table recognition and structure extraction. FinTabNet consists of pages from annual reports of S\&P 500 companies, featuring complex tables.

We used the page images from FinTabNet to generate queries and answers using our automated pipeline, which includes a filtering process. Additionally, we explored using Qwen2-VL-72B for generating this training data, with results reported in \cref{Table:VLM_ablation}. This process resulted in approximately 46k triplets of page images, queries, and answers.

For fine-tuning, we trained both ColPali and ColQwen on our table-focused training data, combined with the original ColPali training set, for one epoch, producing the TabCol models. For RobTab models, we incorporated LLaMA-3-70B for rephrasing, where half of the newly generated training set was randomly rephrased using three different rephrasing levels. These models were then fine-tuned for one epoch, together with the ColPali rephrased dataset.

All fine-tuning procedures followed the same configurations as the official ColPali training pipeline, utilizing ColBERT in-batch loss. The specific configurations for ColPali and ColQwen can be found at \href{https://github.com/illuin-tech/colpali}{GitHub}.  Each model was fine-tuned starting from its respective base version: ColPali from ColPali-1.2 and ColQwen from ColQwen2-1.0. We used a batch size of 64 per GPU, resulting in a total effective batch size of 256. The training maintained the same initial learning rate, warmup steps, learning rate schedule, and LoRA configuration as the original pipeline.  All training runs were conducted on four A100 80GB GPUs, with each fine-tuning session taking approximately three hours to complete.

\subsection{Benchmark Document Examples}  
\begin{figure*}[h!]
    \includegraphics[width=1.1\textwidth]{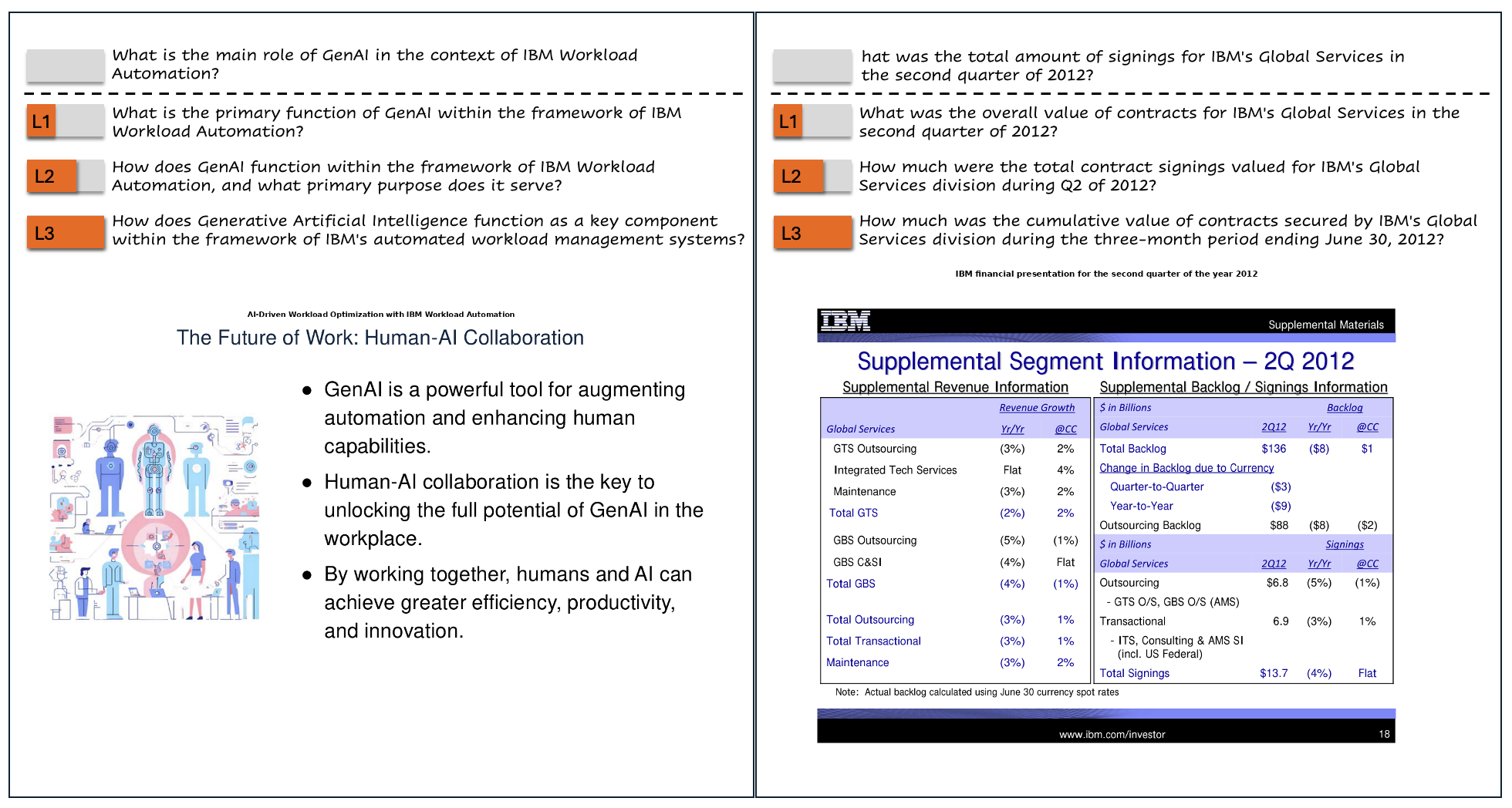} 
    \caption{\textbf{Real-MM-RAG Benchmark Examples with Rephrasing.}  
    On the left: FinSlides—financial quarterly presentations.  
    On the right: TechSlides Technical slides about business and IT automation. 
    Questions are listed from the original query to Level 3 rephrasing.}
    \label{fig:ours_examples_1} 
\end{figure*}

\begin{figure*}[h!]
    \includegraphics[width=1.1\textwidth]{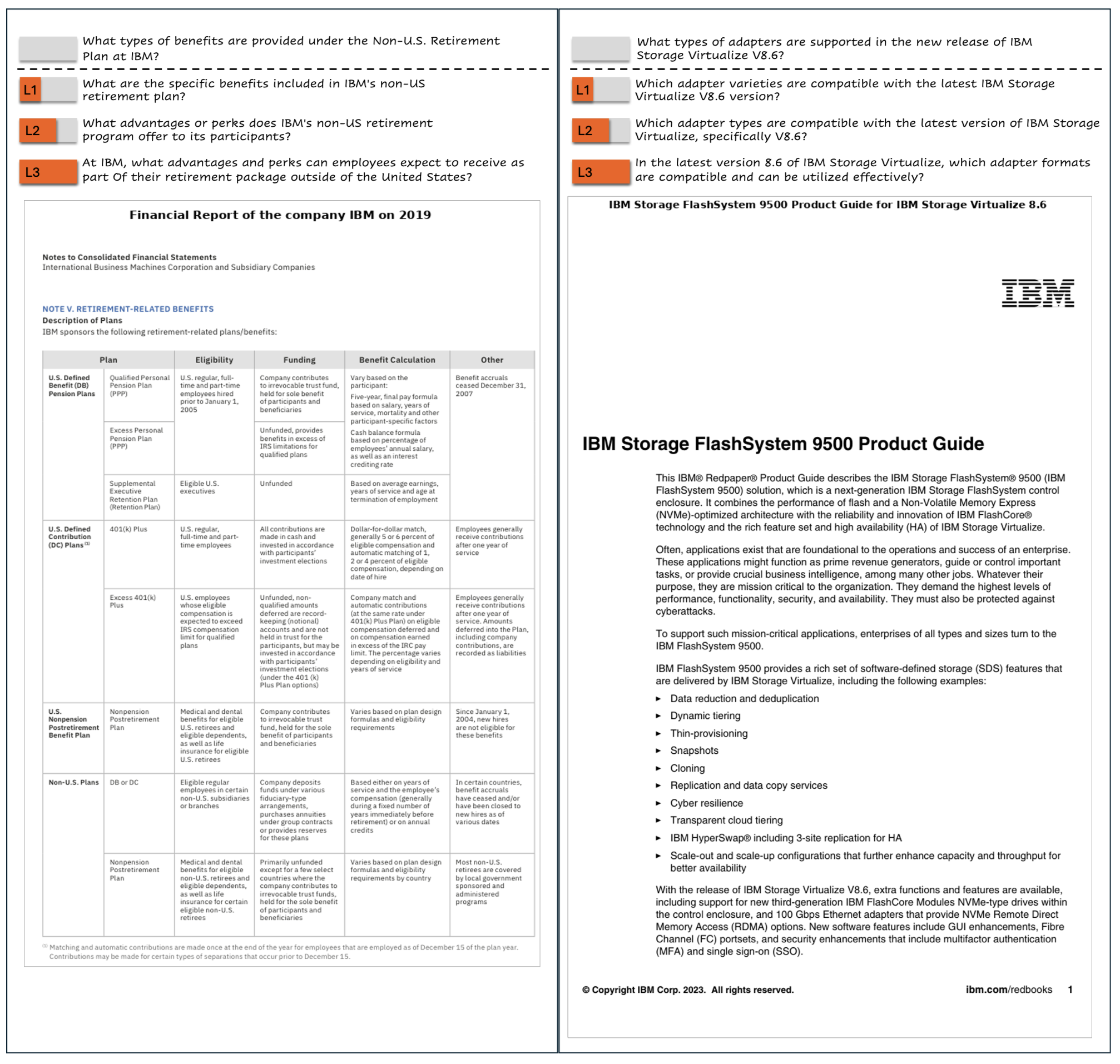} 
    \caption{\textbf{Real-MM-RAG Benchmark Examples with Rephrasing.}  
    Left: FinReport—financial annual reports.  
    Right: TechReport—FlashSystem technical reports.  
    Queries are listed from the original to Level 3 rephrasing.}
    \label{fig:ours_examples_2} 
\end{figure*}

\begin{figure*}[h!]
    \includegraphics[width=1.1\textwidth]{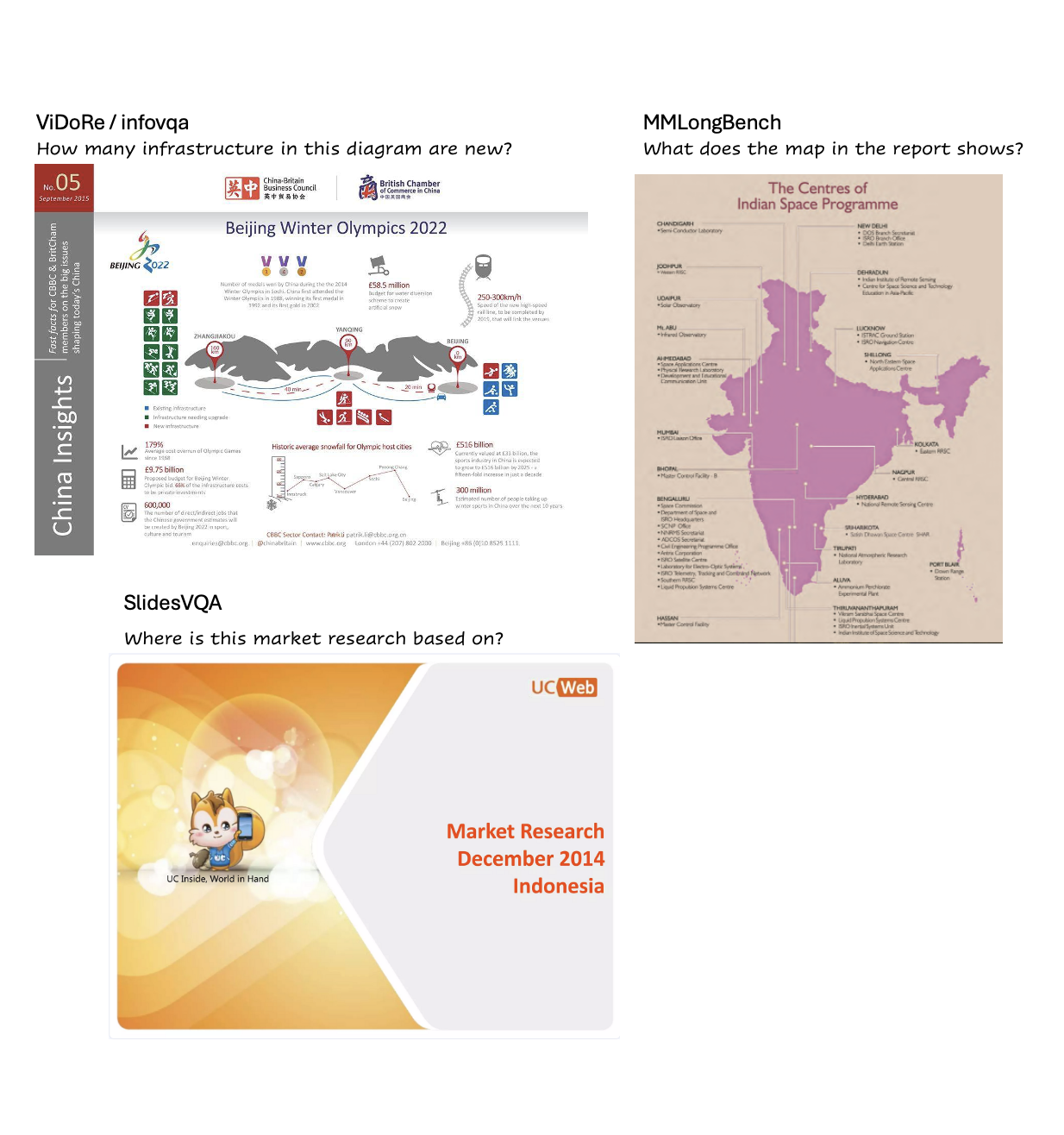} 
    \caption{\textbf{Examples from Previous Benchmarks.} These examples illustrate common query types in these benchmarks. Many queries are generated for question answering and refer to a specific page rather than resembling real user queries, which are typically asked without prior knowledge of a specific page.}

    \label{fig:others_examples} 
\end{figure*}

Our benchmarks include a diverse set of pages containing different types of information, including full-text pages, table-heavy pages, and slides with both tables and other visual elements. \Cref{Table:benchmark_statistics} presents the benchmark statistics for the different datasets.   Additionally, for each page, we provide four types of queries: the original VLM-generated query and three levels of rephrasing. Examples of pages along with all query versions are shown in \cref{fig:ours_examples_1,fig:ours_examples_2}. In \cref{fig:others_examples}, we present examples from previous benchmarks, highlighting that many of their queries are not well-suited for RAG, as they often reference specific pages (i.e., QA-style queries) rather than general information-seeking queries.

\subsection{Additional Results and Ablations} \label{sec:additional_results}
In our main paper, we focused on reporting the NDCG@5 metric for a subset of models and benchmarks across different rephrasing levels. In \cref{Table:rephrasing_benchmarks,Table:rephrasing_benchmarks_recall1,Table:rephrasing_benchmarks_recall5}, we provide the full results for Recall@1, Recall@5, and NDCG@5.  Additionally, \cref{Table:model_comparisons_on_labels} presents model performance across different evidence source types. A Vision-Language Model (VLM), Pixtral-12B, was used to classify each query based on the type of evidence source from which the answer was retrieved on the corresponding page. We analyze performance across three different evidence types: Text, Table, and Visual. The reported results show NDCG@5 scores on the non-rephrased version, averaged across our four benchmarks. These results highlight a significant weakness in handling tables. However, after fine-tuning on the table-focused dataset, we observe improvements across all evidence types, with tables showing the most substantial gains.

We further provide two ablation experiments. The first, shown in \cref{Table:Training_Set_ablation}, aims to demonstrate that the observed improvements after fine-tuning with our proposed datasets are due to the tailored data rather than fine-tuning itself. To verify this, we performed an additional fine-tuning run using the same training set as the RobCol models but without rephrasing, maintaining the exact number of training examples and identical training configurations.  

As seen in the results, ColPali gains some improvement from fine-tuning alone, but the gains are significantly lower compared to fine-tuning with our rephrased training set. This gap is even more pronounced when evaluating on the rephrased benchmark (Level 3). For ColQwen, the baseline fine-tuning without rephrasing leads to a decrease in performance, whereas our fine-tuned models show substantial improvements on rephrased queries, as expected when training with our rephrased dataset.  

The second ablation (see \cref{Table:VLM_ablation}) aims to show that the improvements from the table-heavy training set are general and not specific to the Vision-Language Model (VLM) used for question generation. To test this, we generated an alternative version of the training set using a different VLM, Qwen2-VL-72B-Instruct, and trained models both with and without rephrasing (using LLaMA-3-3-70B). After filtering, the dataset size was slightly smaller—40k examples compared to 46k with Pixtral generation—likely due to Qwen generating more queries that were filtered out.  

While results show a slight decrease in performance compared to using Pixtral-generated data, the fine-tuned models still significantly outperform the base ColPali and ColQwen models. This confirms that the effectiveness of our data is not limited to a specific VLM and that training on a table-heavy dataset remains highly beneficial.

\subsection{Licensing and Additional General Information}  

All models and datasets used in this work comply with their respective licenses. Qwen2-VL (ColQwen2) is licensed under Apache 2.0, with adapters under MIT. PaliGemma (ColPali) follows the Gemma license, with adapters under MIT. Pixtral-12B-2409 (mistralai) and Mixtral-8x22B are both under Apache 2.0, allowing unrestricted use, modification, and distribution. LLaMA 3.3 70B is licensed under the LLaMA 3.3 Community License Agreement. All datasets used are in English. The Colapli training set consists of subsampled academic datasets redistributed under their original licenses. It also includes synthetic datasets generated from publicly available internet data and VLM-generated queries, which are released without usage restrictions. Benchmark datasets are derived from openly available documents and images, with owner approval for publication. Fine-tuning data (Fintabnet) is collected from publicly available sources and processed for compatibility with our models.  

For human evaluation, we published an online form alongside a request for participation in annotating queries for evaluating data intended for publication.
Throughout the paper, an AI assistant (ChatGPT) was used for minor grammar and sentence structure edits.

\begin{table*}[t]
\footnotesize
\renewcommand{\arraystretch}{1.5} 
\setlength\tabcolsep{5pt} 
\caption{
\textbf{Benchmark Statistics.}
}
\vspace{-0.15cm}
\begin{tabular*}{\linewidth}{@{\extracolsep{\fill}}l|ccc c ccc} 
\toprule
\multirow{2}{*}{\textbf{Benchmark}} & \multicolumn{3}{c}{\textbf{Documents}} & \multicolumn{1}{c}{\textbf{Queries}} & \multicolumn{3}{c}{\textbf{Evidence Label}} \\
& \textbf{\# Pages} & \textbf{\# Docs} & \textbf{Avg. Len} & \textbf{\# Queries} &  \textbf{Text} & \textbf{Table} & \textbf{Visual} \\
\midrule
\textbf{FinReport}           & 2687 & 19 & 141  & 853  & 75\% & 24\% & 1\% \\
\textbf{FinSlides}            & 2280 & 65 & 35   & 1052  & 12\% & 83\% & 5\% \\
\textbf{TechReport}       & 1674 & 17 & 98   & 1294  & 81\% & 12\% & 7\% \\
\textbf{TechSlides}      & 1963 & 62  & 32 &  1354 &  66\% & 6\%  & 28\% \\
\bottomrule
\end{tabular*}
\label{Table:benchmark_statistics}
\end{table*}

\hspace{-10cm}

\hspace{-10cm}

\clearpage
\begin{table*}[th]
\footnotesize
\renewcommand{\arraystretch}{1.5} 
\setlength\tabcolsep{4pt} 
\caption{
\textbf{Impact of Rephrasing Levels on Document Retrieval Benchmarks.}
This table shows NDCG@5 performance variations across rephrasing levels (0-3) for different benchmarks and models.}
\begin{tabular*}{1.03\linewidth}{@{\extracolsep{\fill}}l|cccc|cccc|cccc|cccc}
\toprule
 & \multicolumn{4}{c|}{\textbf{FinReport}} & \multicolumn{4}{c|}{\textbf{FinSlides}} & \multicolumn{4}{c|}{\textbf{TechReport}} & \multicolumn{4}{c}{\textbf{TechSlides}} \\
\textbf{Rephrasing Level} & \textbf{0} & \textbf{1} & \textbf{2} & \textbf{3} & \textbf{0} & \textbf{1} & \textbf{2} & \textbf{3} & \textbf{0} & \textbf{1} & \textbf{2} & \textbf{3} & \textbf{0} & \textbf{1} & \textbf{2} & \textbf{3} \\
\midrule

\textit{BM25 (OCR)} & 
48.8 & 38.4 & 26.6 & 21.7 &
13.6 & 13.0 & 7.1 & 5.9 &
66.6 & 48.0 & 38.7 & 35.1 &
58.7 & 45.7 & 35.7 & 31.2 \\

\textit{BGE-M3 (OCR)} & 
47.2 & 40.3 & 37.6 & 36.5 &
16.9 & 16.3 & 14.6 & 11.4 &
44.7 & 40.9 & 37.8 & 37.0 &
60.5 & 56.3 & 51.9 & 49.6  \\

\textit{BM25 (Captioning)} & 
54.4 & 43.0 & 30.6 & 25.3 &
20.9 & 19.0 & 11.0 & 9.9 &
69.0 & 50.8 & 41.8 & 37.2 &
66.4 & 53.4 & 41.7 & 36.1 \\

\textit{BGE-M3 (Captioning)} & 
46.4 & 39.0 & 36.9 & 35.9 &
19.5 & 18.7 & 16.3 & 13.8 &
44.6 & 40.9 & 38.4 & 37.5 &
62.6 & 58.3 & 54.5 & 51.7 \\
\addlinespace
\addlinespace
\textit{ColPali} & 
52.7 & 47.2 & 40.8 & 36.8 &  
62.2 & 59.4 & 37.6 & 27.6 & 
80.6 & 72.9 & 66.5 & 62.0 & 
89.7 & 85.0 & 79.2 & 75.8 \\

\hspace{0.3cm}\textit{RobColPali} & 
59.3 & 57.4 & 51.4 & 47.1 & 
76.8 & 75.1 & 58.3 & 48.4 & 
80.1 & 74.7 & 70.0 & 66.6 & 
90.5 & 88.2 & 84.2 & 82.8 \\

\hspace{0.3cm}\textit{TabColPali} & 
70.5 & 63.5 & 56.4 & 50.5 & 
74.5 & 70.7 & 54.2 & 41.5 & 
82.7 & 73.8 & 66.8 & 61.3 & 
90.8 & 86.5 & 80.4 & 77.6 \\

\hspace{0.3cm}\textit{RobTabColPali} & 
71.0 & 68.5 & \textbf{65.0} & \textbf{63.2} & 
\underline{\textbf{80.9}} & \underline{\textbf{79.5}} & \textbf{68.0} & \textbf{58.3} & 
80.8 & \textbf{76.2} & \textbf{72.6} & \textbf{70.7} & 
90.5 & 87.1 & \textbf{84.3} & \textbf{83.3} \\
\addlinespace
\textit{ColQwen} & 
60.8 & 54.5 & 46.7 & 41.8 & 
59.3 & 54.8 & 39.1 & 31.1 & 
\underline{\textbf{84.2}} & 74.9 & 71.8 & 66.9 & 
91.3 & 85.9 & 80.6 & 78.1 \\

\hspace{0.3cm}\textit{RobColQwen} & 
58.4 & 54.5 & 49.7 & 47.5 & 
65.8 & 63.0 & 52.0 & 44.3 & 
81.9 & 75.4 & 71.8 & 69.5 & 
90.1 & \textbf{87.8} & 84.0 & 83.0 \\

\hspace{0.3cm}\textit{TabColQwen} & 
\textbf{78.2} & \textbf{69.0} & 61.5 & 54.0 & 
77.1 & 73.9 & 58.1 & 49.6 & 
83.6 & 75.1 & 70.8 & 65.9 & 
\textbf{92.4} & 87.7 & 82.5 & 78.9 \\

\hspace{0.3cm}\textit{RobTabColQwen} & 
\textbf{\underline{79.7}} & \textbf{\underline{74.8}} & \textbf{\underline{69.4}} & \textbf{\underline{67.1}} & 
\textbf{79.6} & \textbf{78.5} & \textbf{\underline{69.1}} & \textbf{\underline{61.6}} & 
\textbf{83.7} & \textbf{\underline{79.3}} & \textbf{\underline{75.5}} & \textbf{\underline{73.2}} & 
\textbf{\underline{92.5}} & \textbf{\underline{89.9}} & \textbf{\underline{86.3}} & \textbf{\underline{85.0}} \\

\bottomrule
\end{tabular*}
\label{Table:rephrasing_benchmarks}
\end{table*}

\begin{table*}[t]
\footnotesize
\renewcommand{\arraystretch}{1.5} 
\setlength\tabcolsep{4pt} 
\caption{
\textbf{Impact of Rephrasing Levels on Document Retrieval Benchmarks (Recall@1).}
This table shows Recall@1 performance variations across rephrasing levels (0-3) for different benchmarks and models.}
\begin{tabular*}{1.03\linewidth}{@{\extracolsep{\fill}}l|cccc|cccc|cccc|cccc} 
\toprule
 & \multicolumn{4}{c|}{\textbf{FinReport}} & \multicolumn{4}{c|}{\textbf{FinSlides}} & \multicolumn{4}{c|}{\textbf{TechReport}} & \multicolumn{4}{c}{\textbf{TechSlides}} \\
\textbf{Rephrasing Level} & \textbf{0} & \textbf{1} & \textbf{2} & \textbf{3} & \textbf{0} & \textbf{1} & \textbf{2} & \textbf{3} & \textbf{0} & \textbf{1} & \textbf{2} & \textbf{3} & \textbf{0} & \textbf{1} & \textbf{2} & \textbf{3} \\
\midrule

\textit{BM25 (OCR)} & 
33.9 & 25.4 & 17.1 & 14.0 &
7.7 & 7.5 & 3.7 & 3.0 &
53.4 & 34.1 & 26.1 & 23.1 &
45.9 & 33.4 & 25.4 & 21.3 \\

\textit{BGE-M3 (OCR)} & 
34.0 & 28.6 & 25.8 & 25.0 &
11.0 & 9.5 & 9.3 & 7.2 &
34.4 & 31.2 & 28.6 & 27.8 &
48.6 & 44.2 & 40.0 & 37.7  \\

\textit{BM25 (Captioning)} & 
40.1 & 28.0 & 19.8 & 17.5 &
12.2 & 10.9 & 6.2 & 5.7 &
56.2 & 36.8 & 28.8 & 24.6 &
54.0 & 40.8 & 31.0 & 25.2 \\

\textit{BGE-M3 (Captioning)} & 
33.4 & 27.7 & 25.1 & 25.3 &
12.5 & 11.5 & 9.6 & 8.9 &
34.7 & 31.1 & 28.8 & 27.7 &
50.9 & 45.8 & 41.8 & 40.1 \\
\addlinespace
\textit{ColPali} & 
40.3 & 35.4 & 29.1 & 25.9 &  
45.6 & 41.6 & 23.3 & 15.5 & 
68.1 & 57.6 & 51.5 & 45.8 & 
82.4 & 75.3 & 68.4 & 63.5 \\

\hspace{0.3cm}\textit{RobColPali} & 
44.4 & 42.1 & 36.5 & 32.0 & 
60.4 & 57.4 & 39.8 & 30.4 & 
68.0 & 60.2 & 53.8 & 51.2 & 
82.8 & 79.7 & 74.1 & 73.1 \\

\hspace{0.3cm}\textit{TabColPali} & 
55.1 & 50.3 & 40.9 & 36.0 & 
55.9 & 52.0 & 35.0 & 22.8 & 
70.3 & 58.8 & 50.3 & 44.8 & 
83.9 & 77.8 & 69.4 & 66.5 \\

\hspace{0.3cm}\textit{RobTabColPali} & 
56.5 & 53.9 & 49.4 & 48.4 & 
64.0 & 61.7 & 48.4 & 35.8 & 
67.9 & 61.2 & 57.4 & 54.3 & 
83.4 & 78.4 & 75.2 & 73.1 \\
\addlinespace
\textit{ColQwen} & 
44.0 & 39.6 & 32.6 & 28.5 & 
44.7 & 40.8 & 26.7 & 18.7 & 
73.0 & 60.1 & 56.5 & 51.4 & 
84.2 & 77.4 & 70.3 & 66.5 \\

\hspace{0.3cm}\textit{RobColQwen} & 
41.4 & 37.7 & 34.8 & 33.1 & 
49.1 & 46.6 & 37.2 & 29.3 & 
68.5 & 60.4 & 56.1 & 54.2 & 
83.3 & 79.5 & 74.6 & 72.1 \\

\hspace{0.3cm}\textit{TabColQwen} & 
62.7 & 53.2 & 44.9 & 37.5 & 
58.7 & 55.9 & 41.2 & 33.1 & 
71.0 & 59.8 & 54.9 & 49.7 & 
85.9 & 78.6 & 71.1 & 66.9 \\

\hspace{0.3cm}\textit{RobTabColQwen} & 
58.1 & 52.5 & 50.2 & 45.8 & 
62.2 & 60.7 & 51.4 & 41.8 & 
70.3 & 62.2 & 59.0 & 56.3 & 
85.0 & 80.9 & 76.0 & 74.9 \\

\bottomrule
\end{tabular*}
\label{Table:rephrasing_benchmarks_recall1}
\end{table*}

\begin{table*}[t]
\footnotesize
\renewcommand{\arraystretch}{1.5} 
\setlength\tabcolsep{4pt} 
\caption{
\textbf{Impact of Rephrasing Levels on Document Retrieval Benchmarks (Recall@5).}
This table shows Recall@5 performance variations across rephrasing levels (0-3) for different benchmarks and models.}
\begin{tabular*}{1.03\linewidth}{@{\extracolsep{\fill}}l|cccc|cccc|cccc|cccc} 
\toprule
 & \multicolumn{4}{c|}{\textbf{FinReport}} & \multicolumn{4}{c|}{\textbf{FinSlides}} & \multicolumn{4}{c|}{\textbf{TechReport}} & \multicolumn{4}{c}{\textbf{TechSlides}} \\
\textbf{Rephrasing Level} & \textbf{0} & \textbf{1} & \textbf{2} & \textbf{3} & \textbf{0} & \textbf{1} & \textbf{2} & \textbf{3} & \textbf{0} & \textbf{1} & \textbf{2} & \textbf{3} & \textbf{0} & \textbf{1} & \textbf{2} & \textbf{3} \\

\textit{BM25 (OCR)} & 
62.0 & 49.9 & 35.3 & 29.0 &
19.7 & 18.6 & 10.4 & 8.7 &
77.6 & 59.8 & 50.0 & 45.9 &
70.3 & 57.1 & 45.2 & 40.6 \\

\textit{BGE-M3 (OCR)} & 
58.4 & 50.5 & 47.6 & 47.0 &
22.9 & 22.8 & 19.8 & 15.6 &
53.5 & 49.2 & 45.7 & 45.2 &
70.6 & 66.5 & 62.3 & 60.2  \\

\textit{BM25 (Captioning)} & 
66.9 & 56.3 & 40.4 & 32.5 &
29.4 & 27.2 & 15.5 & 13.8 &
79.7 & 63.5 & 53.5 & 49.0 &
77.2 & 64.8 & 51.3 & 46.1 \\

\textit{BGE-M3 (Captioning)} & 
57.0 & 48.8 & 46.7 & 45.5 &
25.9 & 25.1 & 22.7 & 18.3 &
52.9 & 49.3 & 46.6 & 46.2 &
72.6 & 68.8 & 65.7 & 62.0 \\
\addlinespace
\addlinespace
\textit{ColPali} & 
64.0 & 57.8 & 51.7 & 47.0 &  
76.9 & 74.7 & 50.4 & 38.7 & 
90.7 & 85.7 & 79.4 & 76.1 & 
95.2 & 92.4 & 87.8 & 85.7 \\

\hspace{0.3cm}\textit{RobColPali} & 
73.0 & 71.3 & 64.7 & 60.6 & 
90.6 & 89.8 & 74.1 & 64.4 & 
89.8 & 86.5 & 83.7 & 79.5 & 
96.1 & 94.4 & 91.9 & 90.1 \\

\hspace{0.3cm}\textit{TabColPali} & 
83.8 & 75.5 & 70.1 & 63.3 & 
90.3 & 87.1 & 71.7 & 58.3 & 
92.7 & 86.0 & 80.7 & 75.1 & 
96.0 & 93.2 & 88.9 & 86.7 \\

\hspace{0.3cm}\textit{RobTabColPali} & 
83.1 & 80.8 & 78.0 & 76.0 & 
95.0 & 94.2 & 85.0 & 77.7 & 
91.2 & 88.5 & 85.1 & 84.3 & 
95.6 & 93.7 & 91.2 & 91.3 \\
\addlinespace
\textit{ColQwen} & 
75.3 & 68.0 & 59.6 & 54.0 & 
71.8 & 66.8 & 50.3 & 42.5 & 
93.1 & 86.8 & 84.5 & 79.7 & 
96.4 & 92.3 & 88.9 & 87.4 \\

\hspace{0.3cm}\textit{RobColQwen} & 
73.0 & 69.5 & 63.0 & 60.7 & 
79.8 & 76.5 & 65.6 & 57.8 & 
92.6 & 87.9 & 84.8 & 82.7 & 
95.1 & 94.1 & 91.3 & 91.4 \\

\hspace{0.3cm}\textit{TabColQwen} & 
90.7 & 82.6 & 75.5 & 68.0 & 
92.4 & 88.7 & 72.8 & 64.3 & 
93.3 & 87.2 & 83.8 & 79.6 & 
97.2 & 94.7 & 91.3 & 88.6 \\

\hspace{0.3cm}\textit{RobTabColQwen} & 
86.3 & 80.1 & 76.8 & 74.4 & 
92.5 & 92.1 & 82.0 & 76.2 & 
92.6 & 88.8 & 86.4 & 84.7 & 
96.9 & 95.3 & 93.3 & 92.6 \\

\bottomrule
\end{tabular*}
\label{Table:rephrasing_benchmarks_recall5}
\end{table*}

\begin{table*}[t]
\footnotesize
\renewcommand{\arraystretch}{1.5} 
\setlength\tabcolsep{5pt} 
\caption{
\textbf{Model Performance Across Different Evidence Source Types} 
A Vision-Language Model (VLM) was used to classify each query based on the type of evidence source from which the answer was retrieved on the corresponding page. We present the division of performance across three different source types: Text, Table, and Visual. The reported results are the NDCG@5 scores on the non-rephrased version, averaged across our four benchmarks.}
\begin{tabular*}{\linewidth}{@{\extracolsep{\fill}}lccc} 
\toprule
\textbf{Benchmark} & \textbf{Text} & \textbf{Tables} & \textbf{Visual}  \\
\midrule
\textit{ColPali}  & 75.8 & 58.6 & 84.5    \\
\textit{ColQwen}   & 79.2 & 59.9 & 86.8    \\
\textit{TabColQwen}  & 84.5 & 78.5 & 88.5    \\
\bottomrule
\end{tabular*}
\label{Table:model_comparisons_on_labels}
\end{table*}

\begin{table*}[t]
\footnotesize
\renewcommand{\arraystretch}{1.5} 
\setlength\tabcolsep{4pt} 
\caption{
\textbf{Ablation of Fine-Tuning Without Rephrasing.} To demonstrate that the performance improvement is not solely due to fine-tuning, we fine-tune the models on the original ColPali dataset without rephrasing, using the exact same fine-tuning configuration.}
\centering
\begin{tabular*}{1.03\linewidth}{@{\extracolsep{\fill}}l cc cc cc cc} 
\toprule
 & \multicolumn{2}{c}{\textbf{FinReport}} & \multicolumn{2}{c}{\textbf{FinSlides}} & \multicolumn{2}{c}{\textbf{TechReport}} & \multicolumn{2}{c}{\textbf{TechSlides}} \\
\textbf{Rephrasing Level} & \textbf{0} & \textbf{3} & \textbf{0} & \textbf{3} & \textbf{0} & \textbf{3} & \textbf{0} & \textbf{3} \\

\midrule
ColPali & 
52.7 & 34.5 &  
62.2 & 27.6 & 
80.6 & 62.0 & 
89.7 & 75.8 \\
\hspace{0.3cm}\textit{ColPali Baseline FT} & 
57.1 & 39.1 & 
69.1 & 35.7 & 
80.0 & 61.4 & 
90.0 & 77.8 \\
\hspace{0.3cm}\textit{RobColPali} & 
59.3 & 47.1 & 
76.8 & 48.4 & 
80.1 & 66.6 & 
90.5 & 83.3 \\
\textit{ColQwen} & 
60.8 & 41.8 & 
59.3 & 31.1 & 
84.2 & 66.9 & 
91.3 & 78.1 \\
\hspace{0.3cm}\textit{ColQwen Baseline FT} & 
58.1 & 39.3 & 
53.4 & 27.6 & 
79.7 & 62.1 & 
90.5 & 77.4 \\
\hspace{0.3cm}\textit{RobColQwen} & 
58.4 & 47.5 & 
65.8 & 44.3 & 
81.9 & 69.5 & 
90.1 & 83.0 \\

\bottomrule
\end{tabular*}
\label{Table:Training_Set_ablation}
\end{table*}

\begin{table*}[t]
\footnotesize
\renewcommand{\arraystretch}{1.5} 
\setlength\tabcolsep{4pt} 
\caption{
\textbf{Comparison of different queries generation models.}
This table compares the NDCG5 performance of the ColQwen model fine-tuned with the original data of ColPali and our generated table-focused data from the FinTabNet dataset. The evaluation is conducted for two query generation approaches: one using Pixtral and the other using Qwen. The rephrasing for the benchmarks was performed using LLaMA-3-3-70B. Results are presented across rephrasing levels (0 and 3) for our retrieval benchmarks.}
\centering
\begin{tabular*}{1.03\linewidth}{@{\extracolsep{\fill}}l cc cc cc cc} 
\toprule
 & \multicolumn{2}{c}{\textbf{FinReport}} & \multicolumn{2}{c}{\textbf{FinSlides}} & \multicolumn{2}{c}{\textbf{TechReport}} & \multicolumn{2}{c}{\textbf{TechSlides}} \\
\textbf{Rephrasing Level} & \textbf{0} & \textbf{3} & \textbf{0} & \textbf{3} & \textbf{0} & \textbf{3} & \textbf{0} & \textbf{3} \\

\midrule
ColPali & 
52.7 & 34.5 &  
62.2 & 27.6 & 
80.6 & 62.0 & 
89.7 & 75.8 \\
ColQwen & 
60.8 & 41.8 & 
59.3 & 31.1 & 
84.2 & 66.9 & 
91.3 & 78.1 \\
\midrule
\emph{ColTab (Pixtral Queries Gen.)} & 
78.2 & 54.0 & 
77.1 & 49.6 & 
83.6 & 65.9 & 
92.4 & 78.9 \\

\emph{ColTab (Qwen Queries Gen.)} & 
74.8 & 49.5 & 
74.3 & 41.5 & 
83.8 & 66.8 & 
92.6 & 79.5 \\

\midrule
\emph{\textbf{ColRobTab (Pixtral Queries Gen.)}} & 
79.7 & 67.1 & 
79.6 & 61.6 & 
83.7 & 73.2 & 
92.5 & 85.0 \\

\emph{\textbf{ColRobTab (Qwen Queries Gen.)}} & 
73.5 & 61.1 & 
78.9 & 60.0 & 
82.8 & 71.8 & 
91.8 & 84.7 \\
\bottomrule
\end{tabular*}
\label{Table:VLM_ablation}
\end{table*}

\begin{figure*}[h!]
    \centering
    \includegraphics[width=0.9\textwidth]{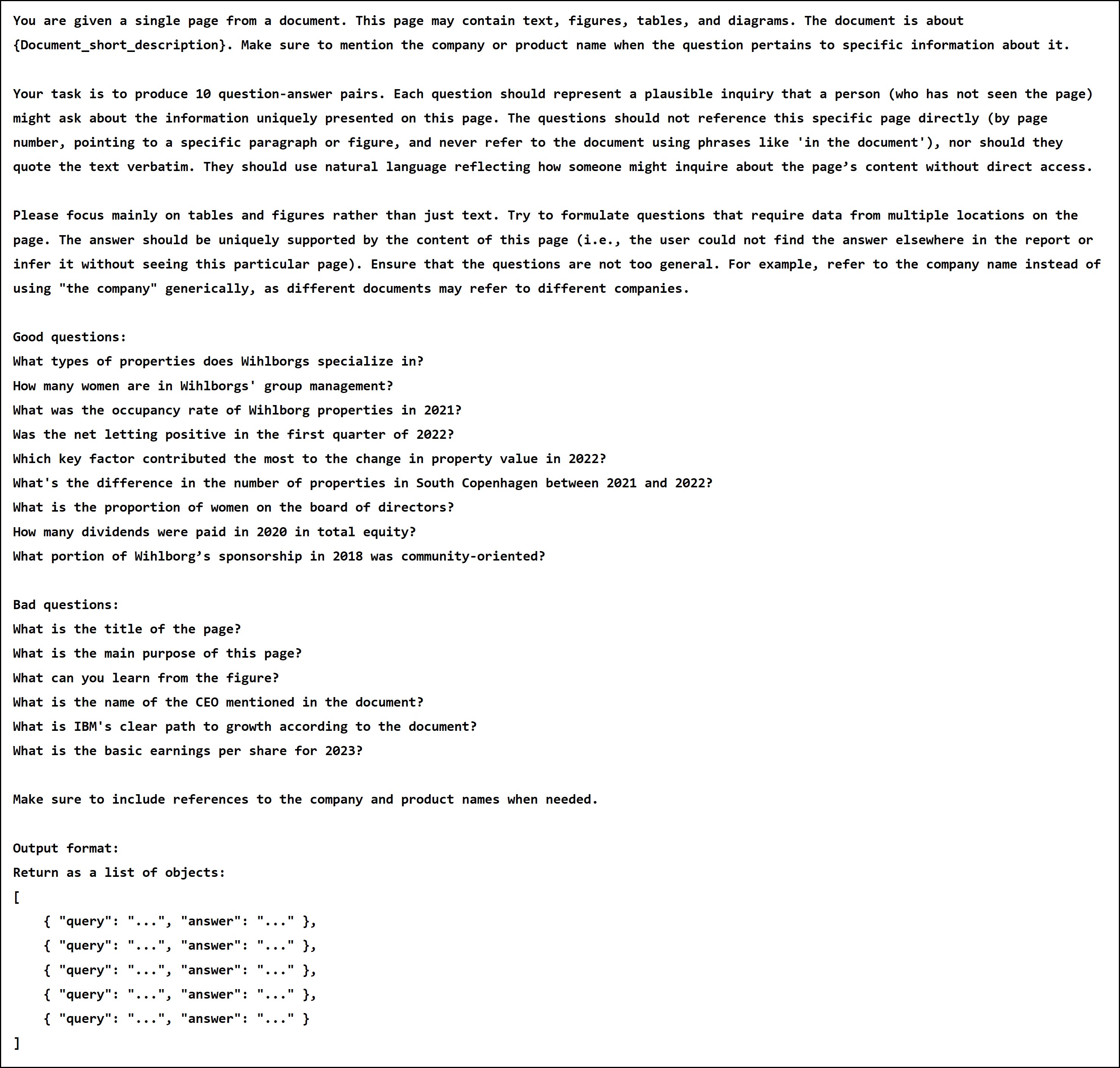} 
    \caption{\textbf{Query Generation Prompt}}
    \label{fig:query_generation_prompt} 
\end{figure*}

\begin{figure*}[ht!]
    \centering
    \includegraphics[width=0.9\textwidth]{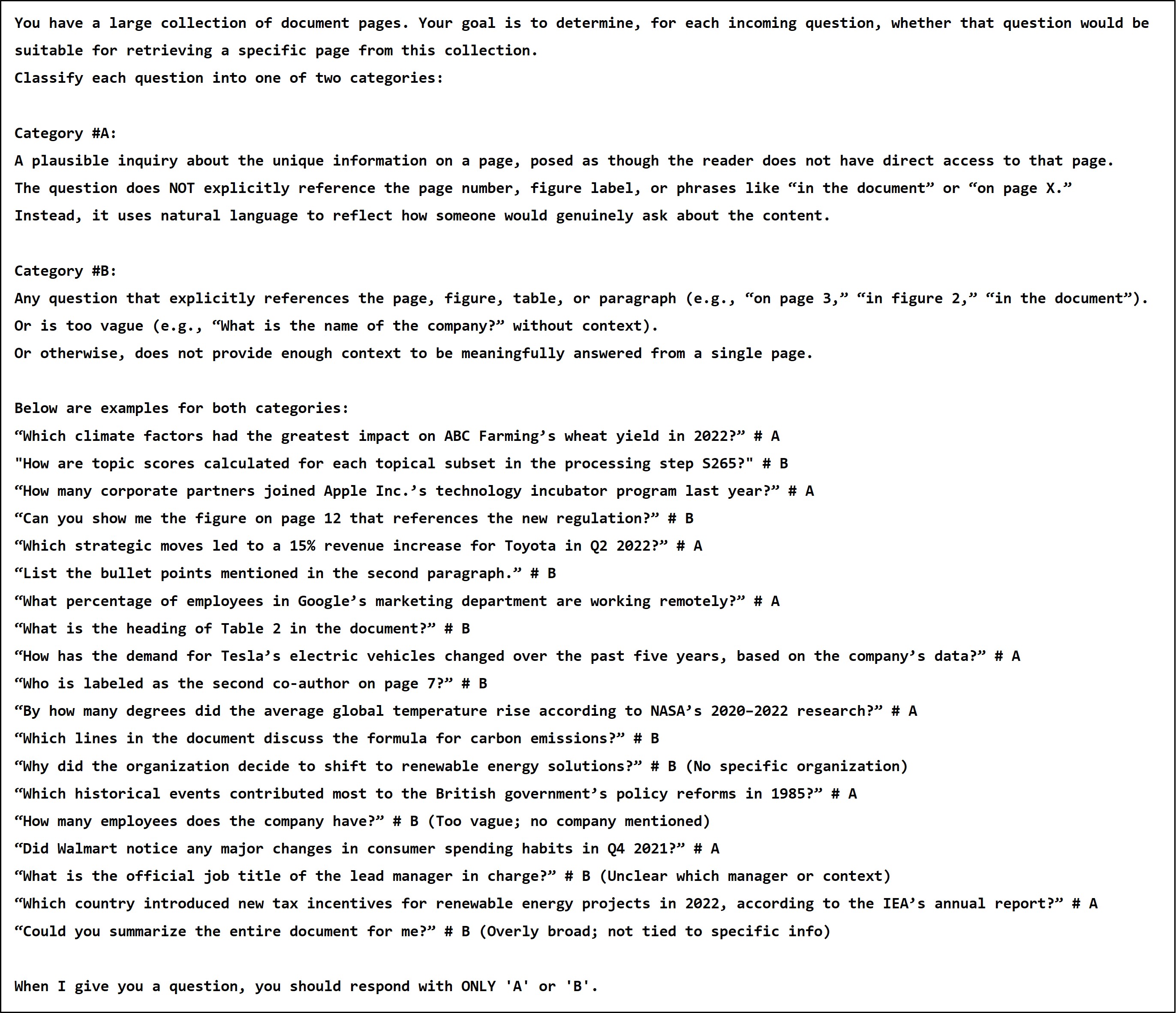}
    \caption{\textbf{Query Verification Prompt}}
    \label{fig:query_verification_prompt} 
\end{figure*}

\begin{figure*}[ht!]
    \centering
    \includegraphics[width=0.9\textwidth]{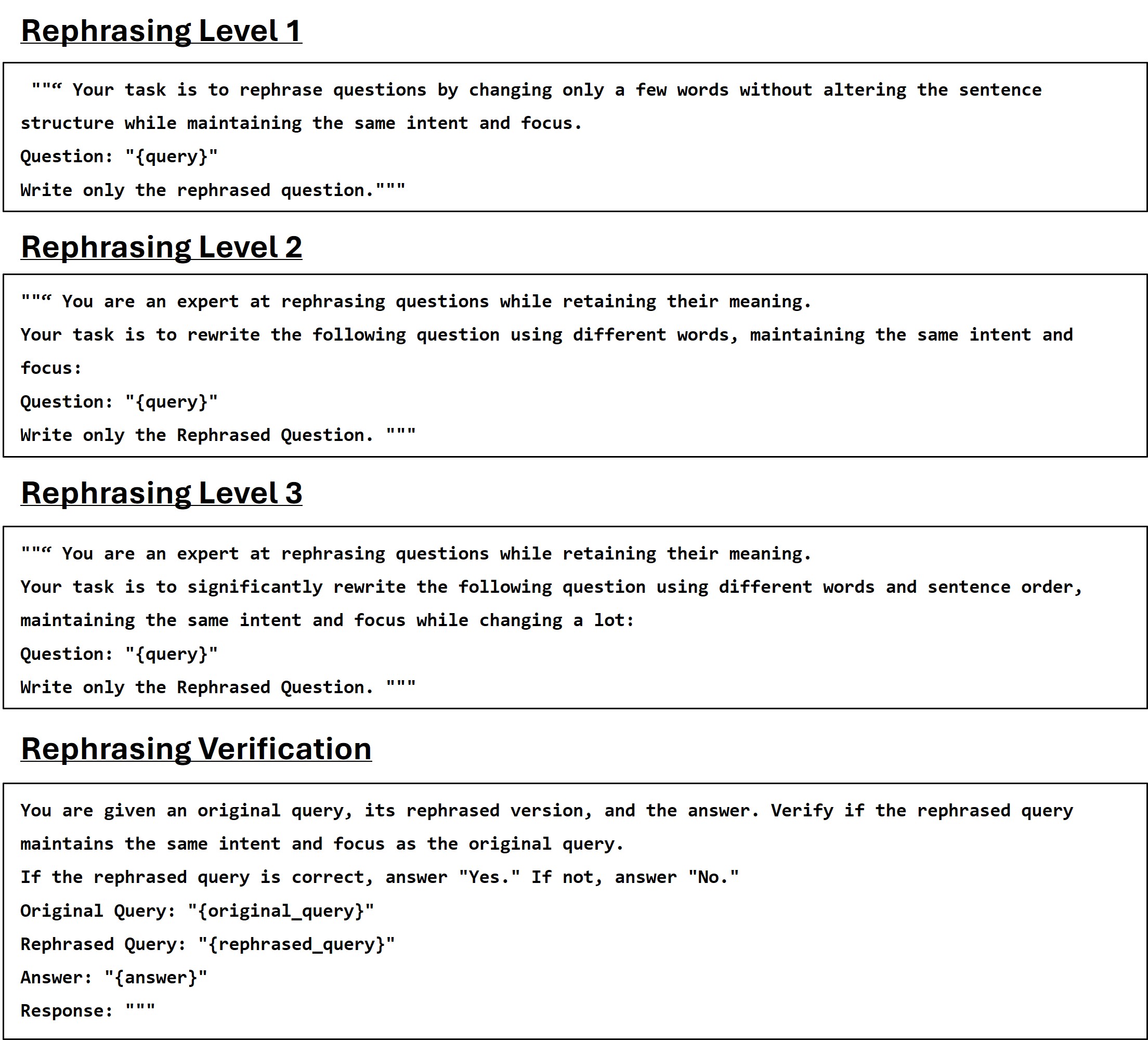}
    \caption{\textbf{Rephrasing Generation and Verification Prompts}}
    \label{fig:rephrasing_prompts} 
\end{figure*}

\begin{figure*}[ht!]
    \centering
    \includegraphics[width=0.9\textwidth]{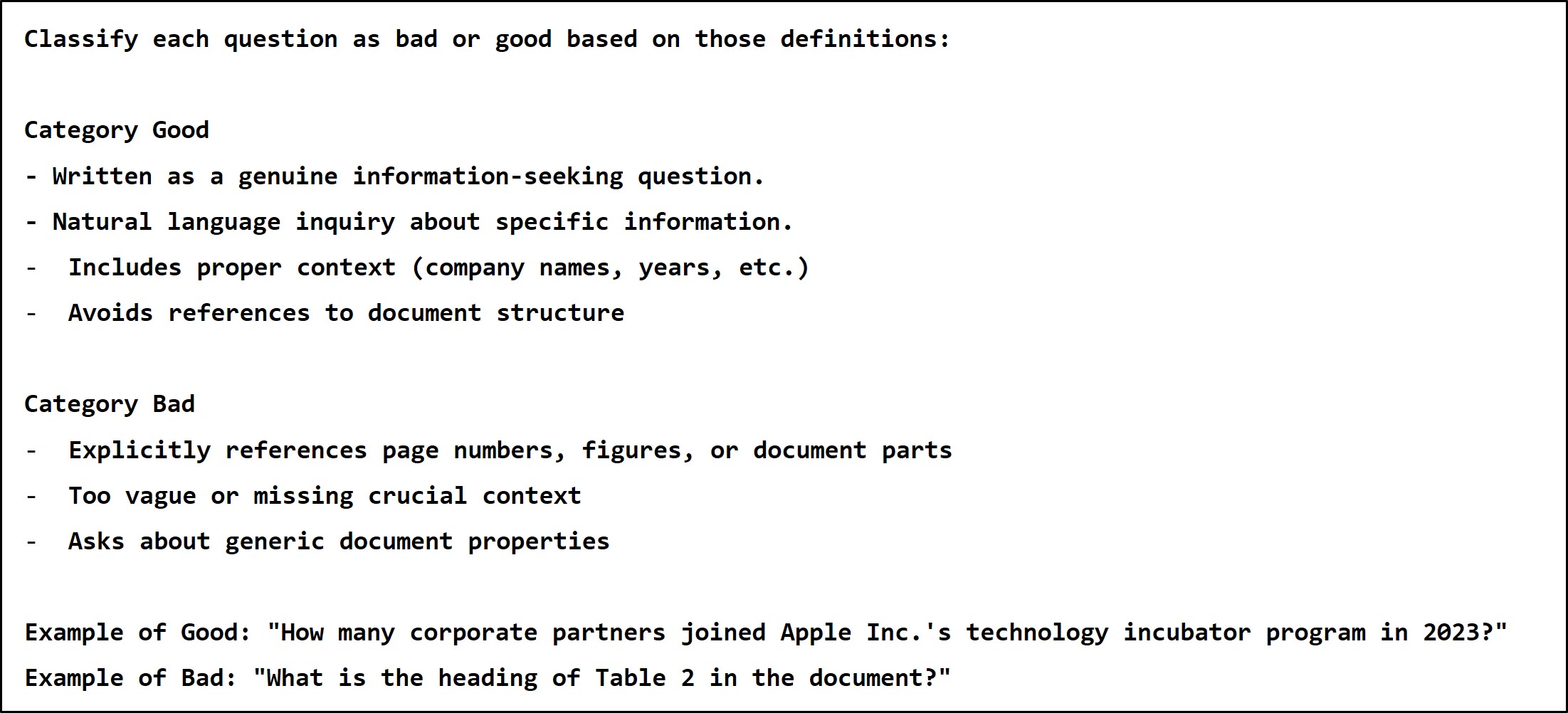}
    \caption{\textbf{Human Evaluation of Query Alignment to RAG.} This figure shows the instructions presented to human annotators along with randomly sampled queries from different benchmarks.}
    \label{fig:Human_query_RAG} 
\end{figure*}

\begin{figure*}[ht!]
    \centering
    \includegraphics[width=0.9\textwidth]{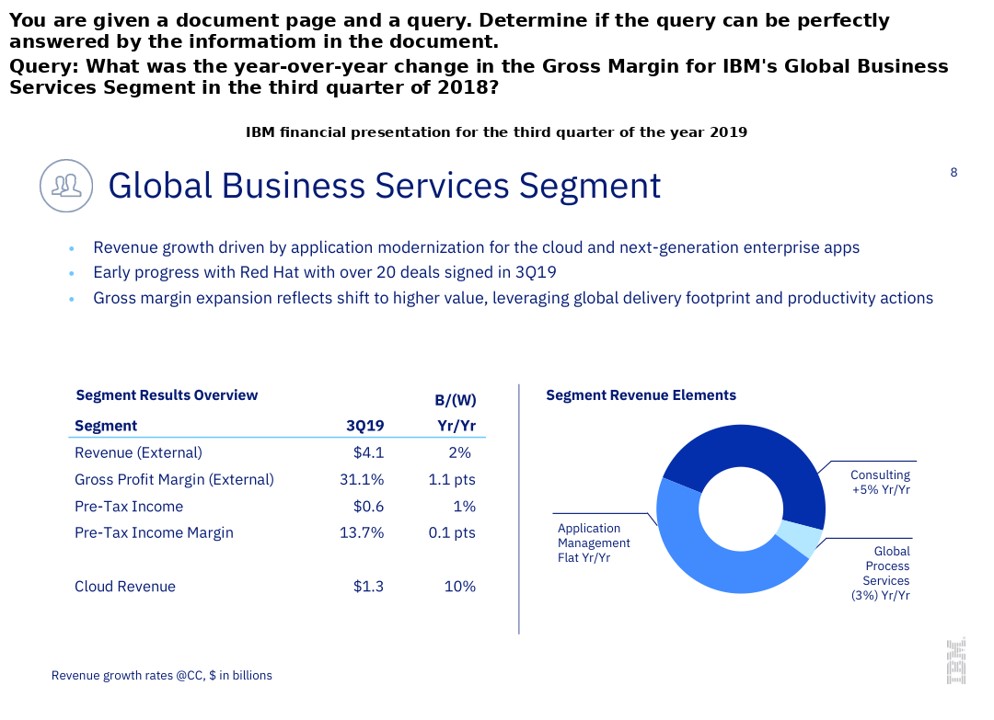} 
    \caption{\textbf{Human Evaluation of False Negatives.} This figure presents an example of the image shown to human annotators, including the instructions, the query, and the negative page retrieved for the given query using ColQwen. The query and page are from our FinSlides benchmark and illustrate a case where the model incorrectly retrieved the wrong year.}

    \label{fig:Human_False_Negative} 
\end{figure*}

\end{document}